%% file: ms.tex
\documentclass[12pt,preprint]{aastex}
\begin{document}
\newcommand{\msun}{\mbox{M$_{\odot}$}}
\newcommand{\rsun}{\mbox{R$_{\odot}$}}
\newcommand{\zsun}{\mbox{Z$_{\odot}$}}
\newcommand{\lsun}{\mbox{L$_{\odot}$}} 

\title{{\it Spitzer} SAGE Infrared Photometry of \\Massive Stars in the
  Large Magellanic Cloud}

\author{A.\,Z. Bonanos\altaffilmark{1}$^{,}$\altaffilmark{2},
D.\,L. Massa\altaffilmark{2}, M. Sewilo\altaffilmark{2},
D.\,J. Lennon\altaffilmark{2},
N. Panagia\altaffilmark{2}$^{,}$\altaffilmark{3},
L.\,J. Smith\altaffilmark{2}, \\ M. Meixner\altaffilmark{2},
B.\,L. Babler\altaffilmark{4}, S. Bracker\altaffilmark{4},
M.\,R. Meade\altaffilmark{4}, K.\,D. Gordon\altaffilmark{2},
J.\,L. Hora\altaffilmark{5}, \\ R. Indebetouw\altaffilmark{6},
B.\,A. Whitney\altaffilmark{7}}

\altaffiltext{1}{Giacconi Fellow.}  \altaffiltext{2}{Space Telescope
Science Institute, 3700 San Martin Drive, Baltimore, MD, 21218, USA;
bonanos, massa, sewilo, lennon, panagia, lsmith, meixner,
kgordon@stsci.edu} \altaffiltext{3}{INAF/Osservatorio Astrofisico di
Catania, Via S.Sofia 78, I-95123 Catania, Italy; and Supernova Ltd., VGV
\#131, Northsound Road, Virgin Gorda, British Virgin Islands.}
\altaffiltext{4}{Department of Astronomy, 475 North Charter St.,
University of Wisconsin, Madison, WI 53706, USA; brian@sal.wisc.edu,
s\_bracker@hotmail.com, meade@sal.wisc.edu}
\altaffiltext{5}{Harvard-Smithsonian Center for Astrophysics, 60 Garden
St., MS 67, Cambridge, MA 02138, USA; jhora@cfa.harvard.edu}
\altaffiltext{6}{Department of Astronomy, University of Virginia, PO Box
3818, Charlottesville, VA 22903, USA; remy@virginia.edu}
\altaffiltext{7}{Space Science Institute, 4750 Walnut St., Suite 205,
Boulder, CO 80301, USA; bwhitney@spacescience.org}

\begin{abstract}

We present a catalog of 1750 massive stars in the Large Magellanic
Cloud, with accurate spectral types compiled from the literature, and a
photometric catalog for a subset of 1268 of these stars, with the goal
of exploring their infrared properties. The photometric catalog consists
of stars with infrared counterparts in the {\it Spitzer} SAGE survey
database, for which we present uniform photometry from $0.3-24$ $\mu$m
in the $UBVIJHK_{s}$+IRAC+MIPS24 bands. The resulting infrared
color--magnitude diagrams illustrate that the supergiant B[e], red
supergiant and luminous blue variable (LBV) stars are among the
brightest infrared point sources in the Large Magellanic Cloud, due to
their intrinsic brightness, and at longer wavelengths, due to dust. We
detect infrared excesses due to free--free emission among $\sim900$ OB
stars, which correlate with luminosity class. We confirm the presence of
dust around 10 supergiant B[e] stars, finding the shape of their
spectral energy distributions (SEDs) to be very similar, in contrast to
the variety of SED shapes among the spectrally variable LBVs. The
similar luminosities of B[e] supergiants ($\log L/L_{\odot}\geq4$) and
the rare, dusty progenitors of the new class of optical transients
(e.g. SN~2008S and NGC~300~OT), plus the fact that dust is present in
both types of objects, suggests a common origin for them. We find the
infrared colors for Wolf-Rayet stars to be independent of spectral type
and their SEDs to be flatter than what models predict. The results of
this study provide the first comprehensive roadmap for interpreting
luminous, massive, resolved stellar populations in nearby galaxies at
infrared wavelengths.

\end{abstract}


\section{Introduction}
\label{section:intro}

The {\it Spitzer Space Telescope} Legacy Survey called ``Surveying the
Agents of a Galaxy's Evolution'' \citep[SAGE;][]{Meixner06} provides an
unprecedented opportunity to investigate the infrared properties of a
large number of massive stars ($\gtrsim\!\!8\,\msun$\footnote{These
stars are massive enough to fuse iron in their cores; they end their
lives as core-collapse supernovae.}) in the Large Magellanic Cloud
(LMC). The SAGE infrared point source catalog has so far enabled studies
of evolved stars \citep{Blum06, Hora08, Srinivasan09}, young stellar
objects \citep{Whitney08}, and variables \citep{Vijh09}; however, the
hot massive star population remains unexplored. The radiation from hot
massive stars at the mid-infrared wavelengths probed by the IRAC
\citep[3.6, 4.5, 5.8, 8.0 $\mu$m;][]{Fazio04} and MIPS \citep[24, 70,
and 160 $\mu$m;][]{Rieke04} cameras onboard {\it Spitzer} consists of:
thermal blackbody emission modified by the atmospheric opacity,
bound--free and free--free emission (``Bremsstrahlung''), which depend
on the density and velocity structure of the stellar wind, and excess
emission if circumstellar dust (or a cool companion) exists.

\citet{Panagia75} and \citet{Wright75} were the first to calculate the
infrared and radio free--free emission from ionized envelopes around hot
massive stars, as a function of the mass-loss rate ($\mbox{\.M}$),
terminal velocity ($v_{\infty}$) of the wind and temperature (T) for the
cases of optically thin and thick winds to explain the radio emission
from hot stars undergoing mass-loss. They found that the flux as a
function of frequency ($\nu$) and distance (d) for an optically thick,
spherically symmetric wind at infrared and radio wavelengths scales as:
$F_{\nu,thick} \propto {\mbox{\.M}}^{4/3} v_{\infty}^{-4/3}
\nu^{0.6}\mbox{d}^{-2} \mbox{T}^{0.1}$, while for an optically thin
wind: $F_{\nu,thin} \propto \mbox{\.M}^{2} {v_o}^{-2} \mbox{R}^{-1} \;
\mbox{T}^{-1.35}\nu^{-2.1}\; B_{\nu}$, where R is the photospheric
radius, $v_{o}$ the initial wind velocity at the surface of the star,
and $B_{\nu}$ is the Planck function. These calculations, following the
shell model of \citet{Gehrz74}, motivated further infrared and radio
observations of Galactic OB stars. The initial studies produced
controversial measurements of infrared excesses \citep[see][]{Sneden78,
Castor83, Abbott84}, whereas unambiguous excesses were detected in the
radio \citep[e.g.][]{Abbott81}. \citet{Leitherer84} detected infrared
excesses in a large number of early-type stars and attributed the
previous non-detections to their different treatment of intrinsic
stellar colors and interstellar reddening. \citet{Bertout85} determined
the velocity law for this large sample, pointing out that excesses can
only be measured reliably in cases of suitably slow (low $v_{\infty}$)
and/or dense (high $\mbox{\.M}$) winds. In principle, mass-loss rates
can be determined from radio observations, which probe the optically
thick wind, e.g. as done for OB-type stars \citep{Abbott81} and
Wolf-Rayet stars \citep{Barlow81, Abbott86}. However, \citet{Moffat94}
presented observational evidence for clumping in stellar winds, which
yields mass-loss rates that are too high when unaccounted for. While
clumping, when not accounted for, can undoubtedly lead to an
overestimation of the mass loss, the radio observations seem to be least
affected by this problem \citep[see e.g.,][]{Puls06}, possibly because
the wind is less clumped at larger distances from the star. On the other
hand, it has been shown that in many cases a substantial fraction of the
radio flux from early-type stars is of non-thermal origin
\citep{Abbott84, DeBecker07}, thus complicating the interpretation of
radio observations.

Until now, work on infrared excesses in hot stars has been limited by
(i) the accuracy of ground based observations at wavelengths greater
than 5 $\mu$m, where infrared excess due to free--free or dust emission
becomes significant, and (ii) the systematic errors associated with
disentangling reddening and distances for Galactic stars, which can be
substantial. The advent of {\it Spitzer}, therefore, provides an
opportunity to readdress and quantify the infrared excesses in hot
massive stars at half-solar metallicity. For the first time, a large
number of stars in the LMC have been observed uniformly out to 24
$\mu$m. The low foreground reddening of this sample of stars, located at
the same distance, circumvents the problems encountered with extincted
Galactic early-type stars, which are observed against a bright infrared
background and distributed at various distances throughout the Galactic
plane. With the goal of characterizing their infrared properties, we
have compiled a catalog of luminous massive stars with known spectral
types in the LMC and extracted infrared photometry for these stars from
the SAGE database. Our results not only serve as a valuable starting
point for more detailed work on specific types of massive stars, but
also provide a roadmap for interpreting existing and future infrared
observations of the most luminous stars in nearby galaxies.

Besides the LMC, {\it Spitzer} has imaged the following Local Group
galaxies: the SMC \citep{Bolatto07}, M31 \citep{Mould08}, M33
\citep{Verley07, McQuinn07, Thompson08}, WLM \citep{Jackson07a}, IC~1613
\citep{Jackson07b}, NGC~6822 \citep{Cannon06}, and the dwarf irregulars:
Phoenix, LGS~3, DDO~210, Leo~A, Pegasus, and Sextans A
\citep{Boyer09}. Only a small fraction of the massive stars in these
galaxies have spectroscopic classifications, rendering identifications
of objects based on their infrared colors uncertain. Our results, which
are based on stars with accurate spectral classifications, will help
interpret these photometric observations. An advantage of studying stars
in the LMC, rather than in more distant Local Group galaxies, is that
blending problems are minimized. For example, the angular resolution at
the IRAC 3.6 $\mu$m band is $1.\arcsec7$, which corresponds to 0.4~pc in
the LMC \citep[at a distance of 48.1 kpc;][]{Macri06}, and 8~pc in M33
\citep[at a distance of 964 kpc;][]{Bonanos06}, while typical OB
associations have sizes of tens of parsecs (but half light radii of
several parsecs), and young massive clusters can have sizes smaller than
a parsec \citep{Clark05}.

The paper is organized as follows: \S\ref{sec:catalog} describes our
spectroscopic and photometric catalogs of massive stars in the LMC,
\S\ref{sec:cmd} presents the resulting color--magnitude diagrams,
\S\ref{sec:colorcolor} the two--color diagrams, and \S\ref{sec:ob} the
infrared excesses detected in OB stars. Sections \S\ref{sec:wr},
\S\ref{sec:lbv}, \S\ref{sec:sgBe}, \S\ref{sec:other} describe the
infrared properties of Wolf-Rayet stars, luminous blue variables,
supergiant B[e] stars, and red supergiants, respectively, and
\S\ref{sec:summary} summarizes our results.

\section{Catalog of Massive Stars in the LMC}
\label{sec:catalog}

We first compiled a catalog of massive stars with known spectral types
in the LMC from the literature. We then cross-matched the stars in the
SAGE database, after incorporating optical and near-infrared photometry
from recent surveys of the LMC. The resulting photometric catalog was
used to study the infrared properties of the stars. In
\S\ref{sec:sptypecatalog} we describe the spectral type catalog compiled
from the literature, in \S\ref{sec:oirsurveys} the existing optical,
near-infrared and mid-infrared surveys of the LMC that were included in
the SAGE database, and in \S\ref{sec:photcatalog} the cross-matching
procedure and the resulting photometric catalog.

\subsection{Spectral Type Catalog}
\label{sec:sptypecatalog}
The largest existing catalogs of OB stars in the LMC were compiled by
\citet[][1273 stars]{Sanduleak70} and \citet[][1822 stars]{Rousseau78};
however the accuracies of the coordinates (arcminute precision) and
spectral types (from objective prism spectra) are not sufficient for a
comparison with the SAGE database. Despite the monumental effort of
Brian Skiff to update coordinates of these stars by manually identifying
each one from the original findercharts \citep[see][]{Sanduleak08},
improved spectral types require new spectroscopy and acquiring them is
an even more difficult task. We therefore compiled a new catalog of
massive stars from the literature, by targeting OB stars having both
accurate coordinates and spectral classifications, making some of these
available for the first time in electronic format. The literature search
resulted in 1750 entries. We do not claim completeness; however, we have
targeted studies of hot massive stars (defined by their spectral types)
in individual clusters and OB associations, as well as studies of
particular types of massive stars. The largest studies included are:
\citet[][new spectral types for 191 OB stars]{Conti86}, \citet[][study
of supergiants\footnote{We have corrected the typographical error in the
name of the B7 Ia+ star Sk~$-67^\circ$~143 given by
\citet{Fitzpatrick91} to the correct name:
Sk~$-69^\circ$~143.}]{Fitzpatrick88,Fitzpatrick91}, \citet[][new
spectral types for 179 OB stars]{Massey95}, \citet[][study of 19 OB
associations]{Massey00}, the VLT-FLAMES survey \citep[][238 stars in N11
and NGC 2004]{Evans06}, the studies of the 30 Doradus region by
\citet{Schild92}, \citet{Parker93} and \citet{Walborn97}, and the
Wolf-Rayet catalog from \citet{Breysacher99}. Note, we have omitted
stars from the crowded R136 cluster. We have added the 6 known luminous
blue variables \citep{Humphreys94}, 10 supergiant B[e] stars
\citep{Zickgraf06}, 20 Be/X-ray binaries \citep{Liu05,Raguzova05} and
early-type eclipsing and spectroscopic binaries. For comparison, we also
included 147 red supergiants from \citet{Massey03b}, some of which were
originally classified by \citet{Humphreys79}. The completeness of the
catalog depends on the spectral type, e.g. it is $\sim3\%$ for the
unevolved O stars in our catalog \citep[out of an estimated total of
6100 unevolved stars with masses $>20\,\msun$ in the LMC;][]{Massey09},
while the Wolf-Rayet catalog \citep{Breysacher99} is thought to be close
to complete.

Table~\ref{tab:catalog} presents our catalog of 1750 massive stars,
listing: the star name, coordinates in degrees (J2000.0), the reference
and corresponding spectral classification, along with comments. The
names of the stars are taken from the corresponding reference and are
sorted by right ascension. The spectral classifications in the catalog
are typically accurate to one spectral type and one luminosity class. We
retained about a dozen stars from the above studies with only
approximate spectral types (e.g. ``early B''). For double entries, we
included the best spectral classification available, usually
corresponding to the most recent reference. We updated coordinates for
all stars with names from the Sanduleak, Brunet, and Parker catalogs
from Brian Skiff's
lists\footnote{ftp://ftp.lowell.edu/pub/bas/starcats/}. Our catalog
contains 427 stars from the Sanduleak catalog, 81 from the Brunet
catalog and 148 from the Parker catalog.

\subsection{Optical and Infrared Surveys of the LMC}
\label{sec:oirsurveys}
Several large optical and infrared photometric catalogs of the LMC have
recently become available, enabling us to obtain accurate photometry for
its massive star population in the wavelength region $0.3-8$~$\mu$m and
in some cases up to $24$~$\mu$m. The optical surveys are: the $UBVR$
catalog of \citet{Massey02} with 180,000 bright stars in an area
covering 14.5 square degrees, the $UBVI$ Magellanic Clouds Photometric
Survey \citep[MCPS;][]{Zaritsky04} including 24 million stars in the
central 64 square degress, and the OGLE III catalog containing $VI$
photometry of 35 million stars covering 40 square degrees
\citep{Udalski08}. The angular resolution for each survey is
$\sim2.\arcsec6$ for the catalog of \citet{Massey02}, $\sim1.\arcsec5$
for MCPS, and $\sim1.\arcsec2$ for OGLE~III.

The existing near-infrared photometric catalogs include: the Two Micron
All Sky Survey \citep[2MASS;][extended by 6X2MASS]{Skrutskie06} and the
targeted IRSF survey \citep{Kato07}, which contains 14.8 million sources
in the central 40 square degrees of the LMC. 2MASS has a pixel scale of
$2.\arcsec0\times2.\arcsec0$, an average seeing of $2.\arcsec5$ and
limiting magnitudes of $J=15.8$, $H=15.1$ and $K_s=14.3$. IRSF has a
pixel scale of $0.\arcsec45$~pixel$^{-1}$, average seeing of
$1.\arcsec3$, $1.\arcsec2$, $1.\arcsec1$ in the $JHK_s$ bands,
respectively, and limiting magnitudes of $J=18.8$, $H=17.8$ and
$K_s=16.6$. In the mid-infrared, the {\it Spitzer} SAGE survey uniformly
imaged the central 7$^{\circ}$ $\times$ 7$^{\circ}$ of the LMC in the
IRAC and MIPS bands on two epochs in 2005, separated by 3 months
\citep{Meixner06}. The survey has recently produced a combined mosaic
catalog of 6.4 million sources. IRAC, with a pixel scale of
$1.\arcsec2$~pixel$^{-1}$, yields an angular resolution of
$1.\arcsec7-2.\arcsec0$ and MIPS at 24~$\mu$m has a resolution of
$6\arcsec$.

Given the variation in the depth, resolution and spatial coverage of
these surveys, we included the available photometry for the massive
stars in our catalog from all the MCPS, OGLE~III, 2MASS, IRSF and SAGE
catalogs. MCPS has incorporated the catalog of \citet{Massey02} for
bright stars common to both catalogs. We note that due to problems with
zero-point, calibration and PSF fitting of bright stars, MCPS photometry
of bright stars should be used with caution. The authors warn that stars
``brighter than 13.5 mag in $B$ or $V$ are prone to substantial
photometric uncertainty'' \citep{Zaritsky04}. Photometry of higher
accuracy, particularly in the optical, exists in the literature for many
of the stars in our catalog; however, it was not included in favor of
uniformity.

\subsection{Photometric Catalog}
\label{sec:photcatalog}

\subsubsection{Matching Procedure}
We used a near-final pre-release version of the SAGE catalog to obtain
near- to mid-infrared photometry for our catalog of 1750 massive
stars. The SAGE catalog includes the IRAC mosaic photometry source list,
extracted from the combined epoch 1 and epoch 2 IRAC images and merged
together with 2MASS and the 2MASS 6X Deep Point Source catalog
\citep[6X2MASS;][]{Cutri04}, and the MIPS 24 $\mu$m epoch 1 catalog. The
IRAC and MIPS 24 $\mu$m catalogs were constructed from the IRAC and MIPS
24 $\mu$m source lists after applying stringent selection criteria and
are highly reliable. These catalogs are subsets of more complete, but
less reliable archives. More details on the data processing and data
products can be found in \citet{Meixner06} and the SAGE Data Description
Document\footnote{http://irsa.ipac.caltech.edu/data/SPITZER/SAGE/doc/}.

We extracted infrared counterparts to the massive stars in our list from
the SAGE IRAC mosaic photometry catalog by performing a conservative
neighbor search with a 1$''$ search radius and selecting the closest
match for each source. This procedure yielded mid-infrared counterparts
for 1316 of the 1750 sources. The IRAC mosaic photometry catalog, MIPS
24 $\mu$m catalog, IRSF, MCPS, and OGLE~III catalogs were
crossed-matched in the SAGE database to provide photometry for sources
over a wavelength range from $0.3-24\,\mu$m.  We used this ``universal
catalog'' to extract multi-wavelength photometry for IRAC sources
matched to the massive stars. Specifically, for IRAC sources with one or
more matches in other catalogs (all but 5), we only considered the
closest matches between sources from any two available catalogs
(IRAC--MIPS24, IRAC--MCPS, MIPS24--MCPS, IRAC--IRSF, MIPS24--IRSF,
IRAC--OGLE~III, MIPS24--OGLE~III), with distances between the matched
sources of $\leq$1$''$. For example, for a match between the IRAC, MIPS
24 $\mu$m and MCPS catalogs (IRAC--MIPS24--MCPS), we applied these
constrains on the IRAC--MIPS24, MIPS24--MCPS, and IRAC--MCPS
matches. These stringent criteria were used to ensure that sources from
individual catalogs for each multi-catalog match refer to the same star;
they reduced the source list to 1262 sources. Six additional massive
stars had matches within 1$''$ in the MIPS 24 $\mu$m Epoch 1 Catalog,
but not in the IRAC Mosaic Photometry Catalog (IRACC).  We supplemented
the missing IRAC data: two sources had counterparts in the IRAC Mosaic
Photometry Archive (IRACA) and 4 MIPS 24 $\mu$m sources had counterparts
in the IRAC Epoch 1 Catalog (IRACCEP1).  These IRAC sources were the
closest matches to the MIPS 24 $\mu$m sources within 1$''$.  Five out of
6 sources also have matches within 1$''$ in the MCPS
catalog. Table~\ref{tab:matchtype} shows the breakdown of the matched
stars to the catalogs, such that 5 stars were only matched to the IRACC,
88 only to the IRACC+IRSF catalogs etc. The above requirements reduced
the source list to 1268 sources\footnote{Preliminary estimates indicate
that matches against the final catalog will not increase the number of
matched stars by more than $1\%$.}. We supplemented the missing IRAC
data for 6 sources: two sources had counterparts in the IRAC mosaic
photometry archive and 4 MIPS 24 $\mu$m sources had counterparts in the
IRAC epoch 1 catalog. We defer the discussion of misidentifications and
blending to \S\ref{sec:other}.

\subsubsection{Catalog Description}
Table~\ref{tab:phot} presents our final matched catalog of 1268 stars,
with the star name, IRAC designation, $UBVIJHK_{s}$+IRAC+MIPS24
photometry and errors, reference paper, corresponding spectral
classification and comments, sorted by increasing right
ascension. Overall, the photometry is presented in order of shortest to
longest wavelength. The 17 columns of photometry are presented in the
following order: $UBVI$ from MCPS, $VI$ form OGLE~III, $JHK_s$ from
2MASS, $JHK_s$ from IRSF, IRAC 3.6, 4.5, 5.8, 8.0 $\mu$m and MIPS 24
$\mu$m. A column with the associated error follows each measurement,
except for the OGLE~III $VI$ photometry. Henceforth, $JHK_s$ magnitudes
refer to 2MASS photometry, whereas IRSF photometry is denoted by a
subscript, e.g. $J_{IRSF}$. All magnitudes are calibrated relative to
Vega \citep[e.g. see][for IRAC]{Reach05}. In Table~\ref{tab:filters}, we
summarize the characteristics of each filter: effective wavelength
$\lambda_{\rm eff}$, zero magnitude flux (in $Jy$), angular resolution,
and the number of detected stars in each filter.

\begin{figure}[ht]  
\includegraphics[width=6in]{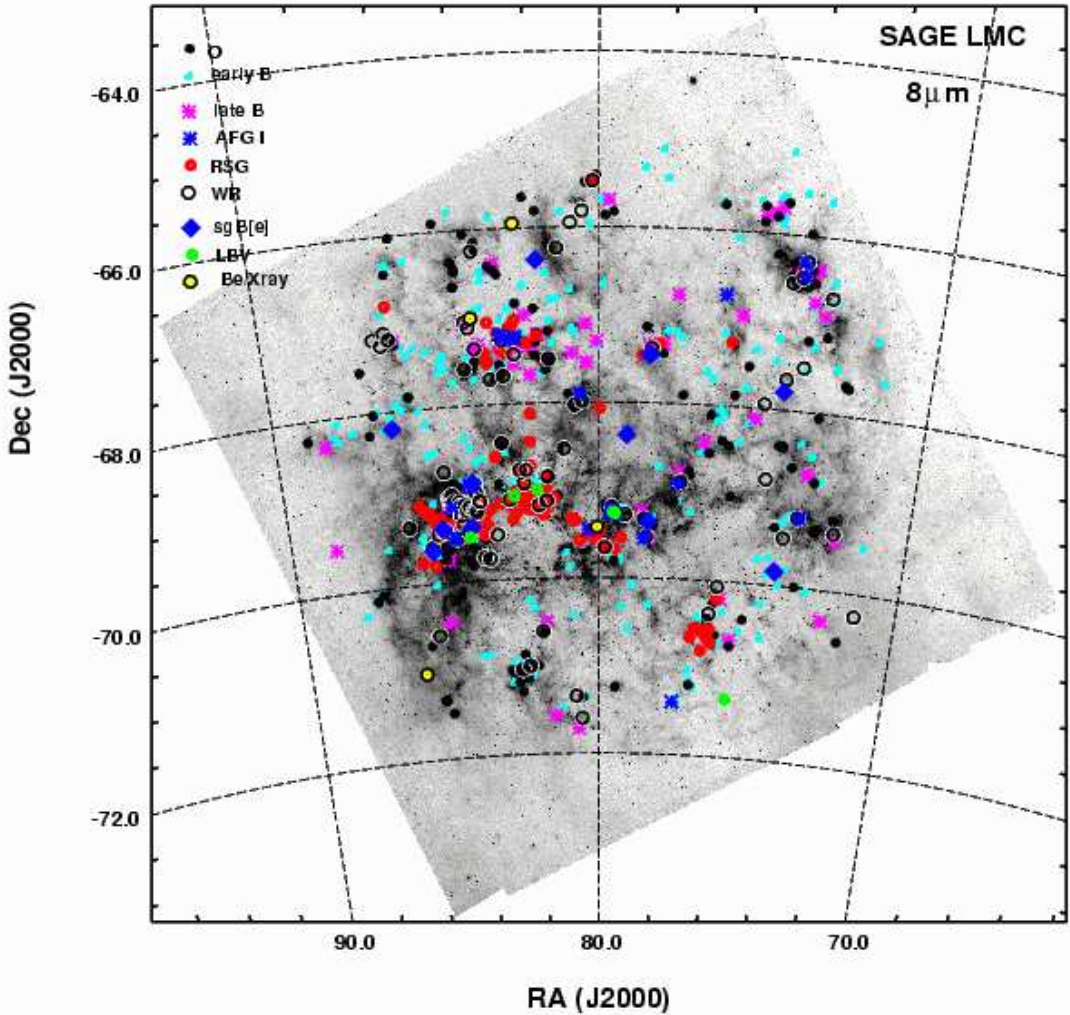}
\caption{Spatial distribution of massive stars with IRAC counterparts,
which are found to trace the spiral features in the 8~$\mu$m SAGE mosaic
image. Different symbols denote different spectral types. The spatial
distribution of the OB stars is uniform, with the exception of the N11
(at $\sim74^{\circ}$, $-66.5^{\circ}$), NGC 2004 (at $\sim82^{\circ}$,
$-67.3^{\circ}$) and 30 Doradus (at $\sim85^{\circ}$, $-69^{\circ}$)
regions.}
\label{spatial}
\end{figure}

The spatial distribution of our 1268 matched sources is shown in
Figure~\ref{spatial}, overlayed onto the 8~$\mu$m image of the LMC. We
find that the massive star population traces the spiral features of the
8~$\mu$m emission, which maps the surface density of the interstellar
medium. Despite the fact that most studies we included from the
literature targeted individual clusters, the spatial distribution of OB
stars in our catalog is fairly uniform, with the exception of the N11,
NGC 2004 and 30 Doradus regions, which have been subjects of several
spectroscopic studies.

\section{Color--Magnitude Diagrams}
\label{sec:cmd}

\begin{figure}[ht]  
\plotone{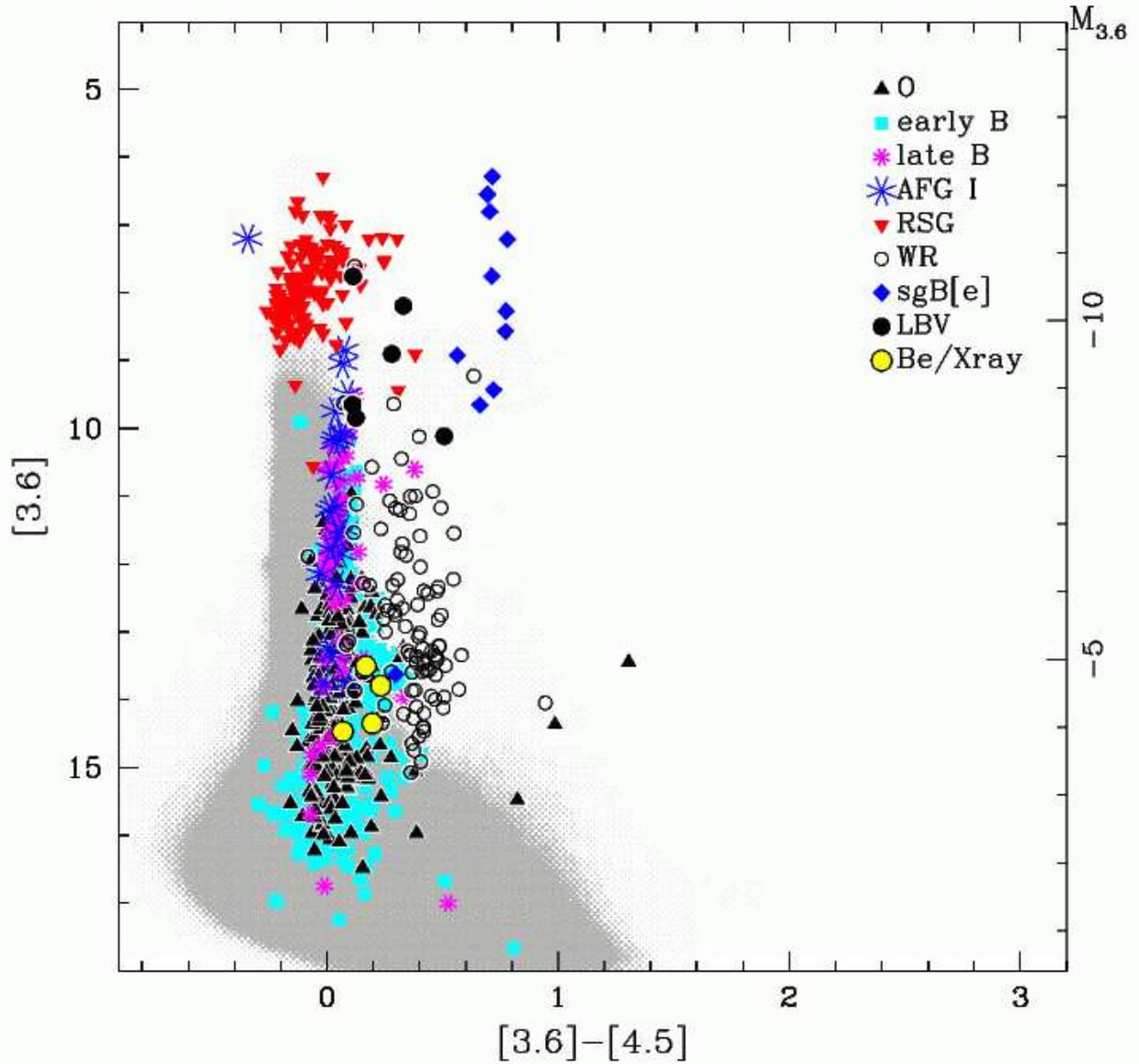}
\caption{[3.6] vs. $[3.6]-[4.5]$ color magnitude diagram for massive
stars with IRAC counterparts in the SAGE database. The conversion to
absolute magnitudes is based on a true LMC distance modulus of 18.41 mag
\citep{Macri06}. Different symbols denote different spectral types. The
locations of all the SAGE detections are shown in grey as a Hess
diagram. The sgB[e], RSG and LBVs are among the most luminous stars at
3.6~$\mu$m.}
\label{cmd36}
\end{figure}

\begin{figure}[ht]  
\plotone{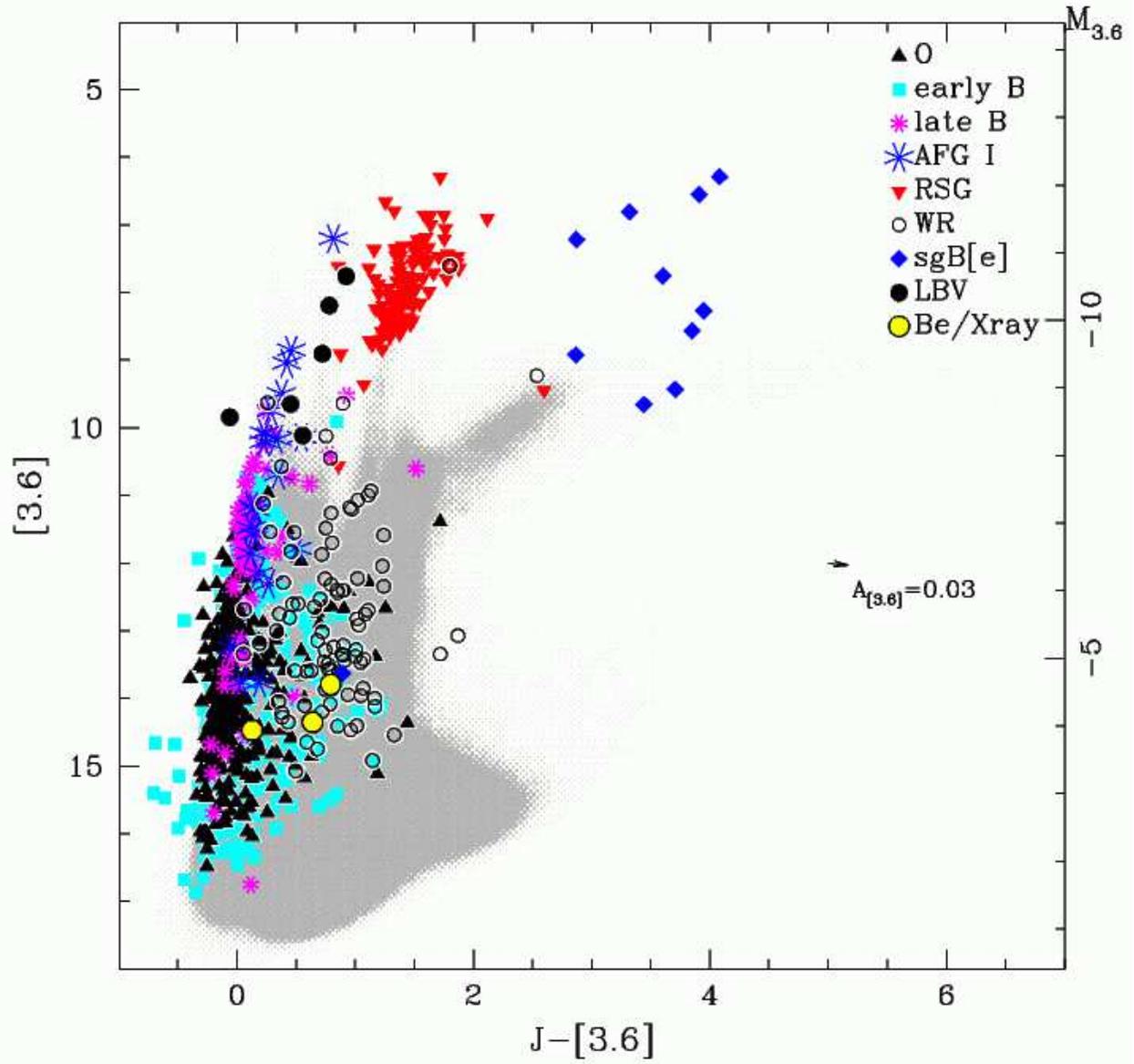}
\caption{Same as Figure~\ref{cmd36}, but for the $[3.6]-J$ vs. [3.6]
color magnitude diagram. The longer baseline separates the populations
more clearly. The reddening vector for $E(B-V)=0.2$ mag is shown.}
\label{cmd36j36}
\end{figure}

We divide the matched stars into 9 categories according to their
spectral types: O stars, early (B0-B2.5) and late (B3-B9) B stars (the
latter have supergiant or giant luminosity classifications, except for 2
B3V stars), spectral type A, F and G (AFG) supergiants, K and M red
supergiants (RSG), Wolf-Rayet stars (WR), supergiant B[e] stars
(sgB[e]), confirmed luminous blue variables (LBV) and Be/X-ray
binaries. In Figures~\ref{cmd36} and \ref{cmd36j36}, we present infrared
$[3.6]$ vs. $[3.6]-[4.5]$ and $J-[3.6]$ color--magnitude diagrams (CMDs)
for all the stars in the catalog, identifying stars in these 9
categories. The conversion to absolute magnitudes in all CMDs is based
on a true LMC distance modulus of 18.41 mag \citep{Macri06}. The
locations of all the SAGE catalog detections are represented by the grey
two dimensional histogram (Hess diagram). The red giants form the clump
at $[3.6]>15$ mag, while the vertical blue extension contains late-type
LMC and foreground stars (free--free emission causes the OB stars to
have redder colors). The asymptotic giant branch (AGB) stars are located
at $[3.6]\sim10$~mag, $J-[3.6]\sim2.5$~mag. Immediately, one notices
that the sgB[e], RSG, and LBVs stars are among the brightest stars at
3.6 $\mu$m and occupy distinct regions in the diagrams. Most of the O
and B stars are located along a vertical line at $[3.6]-[4.5]\sim0$, as
expected. The RSG have ``blue'' colors because of the depression at
[4.5] due to the CO band \citep[see e.g.][]{Verhoelst09}. The WR stars
are on average 0.3 mag redder than the OB stars because of their higher
wind densities and, therefore, stronger free--free emission, while the
sgB[e] stars are 0.6-0.7 mag redder and have similar brightnesses to the
RSG.  The Be/X-ray binaries are located among the ``red'' early-B
stars. The LBVs have similar colors to the WRs, but are brighter. The
late-B stars are brighter than the early-B stars because most of the
former are luminous supergiants. The brightest of the AFG supergiants is
the (only) G-type supergiant Sk~$-69^{\circ}30$. A reddening vector for
$E(B-V)=0.2$ mag is shown in Figure~\ref{cmd36j36} to illustrate the
small effect of reddening, which decreases at longer wavelengths.

In Figures~\ref{cmd8} and \ref{cmd24}, we present $[8.0]$ and $[24]$
vs. $[8.0]-[24]$ CMDs, respectively. At these wavelengths, only stars
with dust are detected, as the sensitivity of {\it Spitzer} drops
sharply, while the flux from hot stellar photospheres also
decreases. Therefore only the dusty WR, LBVs, sgB[e], and RSG are
detected, as well as certain OB stars and late B and AFG supergiants,
which are likely associated with hot spots in the interstellar medium,
and may well be interacting with it. See \S\ref{sec:wr}, \ref{sec:lbv},
\ref{sec:sgBe}, and \ref{sec:other} for a more detailed description of
their colors. Stars with cool dust are brightest at 24~$\mu$m, while
emission from warmer dust peaks at 8~$\mu$m. In these CMDs, the
locations of the massive stars overlap with those of AGB stars, young
stellar objects and planetary nebulae. Infrared colors alone are not
sufficient to distinguish among these very different types of stars,
illustrating the important diagnostic value of wide-baseline photometry.

\section{Two--Color Diagrams}
\label{sec:colorcolor}

In Figures~\ref{jkk80}, \ref{iraccc} and \ref{iraccc2} we present
two--color diagrams (TCDs) using the near and mid-infrared photometry
from our matched catalog. We label stars according to their spectral
types and overplot all the SAGE detections in grey as Hess diagrams. We
find that the majority of the OB stars have colors near zero in all
bands, as expected, while there is an almost continuous extension of OB
star ``outliers'' toward the WR stars, due to thick winds or the
presence of disks. Among these outliers are the Be/X-ray binaries and
other emission line stars. The most conspicuous group of stars in all
TCDs are the B[e] supergiants, which have large excesses, of $\sim4$~mag
in the $K-[8.0]$ color. The RSG occupy a distinct position in the CMDs,
because of their cooler temperatures, but are found to have a range of
mid-infrared colors due to the amount of dust they contain (see
\S\ref{sec:other}). We have overplotted 6 simple theoretical models to
guide the interpretation of the stars in these diagrams. These models
are:

(i) Blackbodies (BBs) at 30,000, 10,000, 5,000, 3,000, 1,700, 1,100, 800
and 500~K.  BBs are a good approximation of stellar emission in the
infrared for most stellar atmosphere spectral energy distributions
(SEDs), with the possible exception of red supergiants, which have
spectra dominated by emission lines.

(ii) A power law model F$_\nu \propto \nu^{\alpha}$, for $\alpha$
ranging from $-1.5$ to 2 in steps of 0.5.  In the infrared, most stellar
SEDs are close to the Rayleigh-Jeans tail of a BB and correspond to
$\alpha\simeq 2$.  As clearly illustrated by \citet{Panagia75} and
\citet{Wright75} \citep[see also][]{Panagia91}, ionized stellar winds
display flatter SEDs than BBs.  This is because at low optical depths
the gas emission in the wind (mostly free--free emission) is rather
flat, and in the optically thick regime the increase of the size of the
emitting region as a function of wavelength (e.g. $R\propto
\lambda^{0.7}$ for a constant velocity wind) partially compensates for
the intrinsic thermal $\lambda^{-2}$ behaviour. Wind SED indexes
asymptotically reach $\alpha=0.6$ in the case in which {\it both} the
expansion velocity is constant (e.g. at suitably large radii where the
terminal velocity has been attained) {\it and} the optical depth is high
(e.g. long wavelengths).

(iii) A model comprising a 30,000~K blackbody and an ionized wind.  This
approximates the case of all early-type stars (essentially OB and WR
types) whose winds add a flatter spectral component that is noticed as
an excess which becomes more conspicuous at longer wavelengths.
Calculations were made following \citet{Panagia91} prescriptions for a
spherically symmetric wind whose expansion velocity increases with
radius like a power law, $v=v_o(r/r_o)^\gamma$, and for a wide range of
mass-loss rates. Although these power laws represent just a zero-order
approximation to the actual velocity structure of the wind (most
noticeably, lacking a limiting/terminal velocity behaviour as suggested
by ultraviolet line studies), they are able to reproduce the power-law
behaviour of the wind SEDs in the regime of high optical depths, i.e.
where the wind contribution becomes dominant. For the sake of
simplicity, we show here only the curves corresponding to $\gamma=0.5$
and $\gamma=0$. Since the gas opacity is an increasing function of the
wavelength, for each given mass-loss rate, the wind emission is higher
at long wavelengths.  As a result the corresponding trajectories on the
various TCDs are curves that start at the colors appropriate for a pure
stellar SED (a power law $\alpha\simeq2$), end at the asymptotic slope
for very optically thick winds (e.g., 0.95 for $\gamma=0.5$ and 0.6 for
$\gamma=0$), and display a characteristic upward concave curvature
because of a more prompt increase in the redder colors (i.e. a positive
second derivative). This is the only model not labeled in
Figures~\ref{iraccc} and~\ref{iraccc2}, due to space constraints.

(iv) A model comprising a 30,000~K blackbody plus emission from an
optically thin ionized nebula.  Since the angular resolution of {\it
Spitzer} is never better than $1.\arcsec7$, which at the LMC distance
corresponds to more than 0.4~pc, we have to consider the case of an
unresolved HII region surrounding an early-type star.  For these models
we adopted the gas emissivity as computed by \citet{Natta76}, which
includes emission lines, bound--free and free--free continuum for both H
and He, and is almost constant over the entire wavelength range
$1-20\,\mu$m. As expected, the curves start from the bare atmosphere
point and asymptotically tend to a point corresponding to a spectral
slope of $\alpha\sim0$, displaying an upward concave curvature.

(v) A model with a 30,000~K blackbody plus cool dust emission (140~K and
200~K BB) with the amount of dust increasing in logarithmic steps of
0.5.  This model may illustrate the case of an early-type star
surrounded by a cool dusty envelope, perhaps similar to a sgB[e] star
\citep{Kastner06}, or stars with cold dust in their vicinity. In these
calculations the dust opacity is assumed to be inversely proportional to
the wavelength ($\kappa_{dust}\propto \lambda ^{-1}$). At these dust
temperatures the peak of the dust emission occurs at about 18 and
26~$\mu$m, and, therefore, the corresponding dust SED is rapidly (almost
exponentially) increasing toward longer wavelengths, thus displaying a
very negative slope ($\alpha<<0$).

(vi) A model with a 3,500~K blackbody plus 250~K dust, which was
selected to represent the case of red supergiants \citep[according to
the new RSG effective temperature scale for a M3~I in the LMC,
see][]{Levesque07}.  Also in these calculations the dust opacity is
assumed to be inversely proportional to the wavelength
($\kappa_{dust}\propto \lambda ^{-1}$), and the amount of dust is
increased in logarithmic steps of 0.5. As expected, the starting point
corresponds to the bare stellar atmosphere and the ending point is where
the dust emission dominates (the peak of the dust emission SED occurs at
about 15~$\mu$m), and the tracks display a typical upward concave
curvature.

From these simple models and considerations, it appears that ionized gas
emission produces a SED that declines with wavelength less steeply than
a blackbody, thus possibly causing an excess both in the infrared and in
the radio domains but without reversing the overall declining trend of
the SED. It follows that an observed flux increase toward longer
wavelengths {\it necessarily} requires the presence of an additional
component that has a suitably low temperature (e.g. less than 500~K for
an upturn around 5~$\mu m$) and has a suitably high emission to affect,
and eventually dominate the observed SED.  These properties correspond
to the characteristics of a dusty layer, that is able to absorb mostly
optical and ultraviolet photons and reradiate that energy in the
infrared.

The locations of WR, LBV, sgB[e] and RSG are discussed in
\S\ref{sec:wr}, \ref{sec:lbv}, \ref{sec:sgBe} and \ref{sec:other},
respectively.

\section{Infrared Excesses in OB stars}
\label{sec:ob}

The primary cause of mid-infrared excess in OB stars is free--free
emission from winds or disks. Since this emission is a monotonically
increasing function of the optical depth that depends on the square of
the electron density, one may find a different mid-infrared emission
from stars with identical mass-loss rates but different wind velocity
fields. Therefore, winds that are slow, clumped
\citep[e.g.,][]{Blomme03} or compressed toward the equatorial plane to
form a disk \citep{Owocki94} will have stronger mid-infrared
emission. Given the sensitivity to these parameters, \citet{Castor83}
and \citet{Abbott84} concluded that mid-infrared emission alone cannot
be used to determine the mass-loss rates of OB stars. More recently,
\citet{Puls06} argued that this very sensitivity makes mid-infrared
emission extremely important in disentangling the nature of clumped
winds. We therefore proceed to examine the mid-infrared excesses of the
354 O and 586 early-B stars in our photometric catalog.

In Figures~\ref{fig:oexcess} and~\ref{fig:bexcess}, we plot $J_{IRSF}$
vs.  $J_{IRSF}-[3.6]$, $J_{IRSF}-[5.8]$ and $J_{IRSF}-[8.0]$ colors for
O and early-B stars, respectively, indicating their luminosity classes,
binarity and emission line classification properties by different
symbols. We compare the observed colors with colors of plane-parallel
non-LTE TLUSTY stellar atmosphere models \citep{Lanz03, Lanz07} of
appropriate metallicity and effective temperatures. We have not
dereddened the stars, because this requires $BV$ photometry of higher
accuracy than that provided by the MCPS catalog; however, the low
extinction toward the LMC \citep[$\overline{E(B-V)}=0.14$
mag;][]{Nikolaev04} makes such corrections small, as illustrated by the
reddening vectors. For reference, TLUSTY models reddened by $E(B-V)=0.2$
mag are also shown. We clearly detect infrared excesses despite not
having dereddened the stars. At longer wavelengths, the excess is larger
because the flux due to free--free emission for optically thin winds
remains essentially constant with wavelength. Fewer stars are detected
at longer wavelengths primarily because of the decreasing sensitivity of
{\it Spitzer} and the overall decline of the SED.

We find that for $J<13$~mag, more luminous OB stars exhibit larger
infrared excesses in all colors due to winds. This can easily be
understood considering that the infrared excess is an increasing
function of the infrared optical depth, which in turn is proportional to
the ratio $\dot M^2/(v_{exp}^2R_*^3)$ \citep[e.g.][]{Panagia91}.  In
early-type stars the mass-loss rate is found to be proportional to a
power $k$ (higher than unity) of the stellar luminosity, i.e. $\dot
M\propto L^k$ with $k\simeq1.2-1.6$ \citep{Lamers93, Scuderi98}. The
average expansion velocity is a function of the effective temperature,
ranging from an initial value at the photosphere $v_0\simeq
v_{sound}\propto T_{\rm eff}^{1/2}$ to a terminal value $v_\infty\propto
T_{\rm eff}^2$ \citep{Panagia82}, so that approximately $v_{exp} \simeq
(v_0 v_\infty)^{0.5} \propto T_{\rm eff}^{1.25}$.  Since the stellar
radius can be expressed as $R\propto L^{1/2}/T_{\rm eff}^2$, it follows
that the infrared optical depth in early-type star winds is proportional
to $L^{(2k-1.5)}\times T_{\rm eff}^{3.5}$, i.e. a $\sim 0.9-1.7$ power
of the luminosity. This result can straightforwardly account for the
observed increase of the infrared excess as a function of luminosity for
O and B type stars in the LMC. We would like to mention that our
findings confirm and provide a simple explanation for the luminosity
effect between hypergiants and dwarfs noted by \citet{Leitherer84} in
their sample of 82 Galactic OB(A) stars. The brightest stars with the
largest excesses shown in Figure~\ref{fig:oexcess} are late O
supergiants, with large $\mbox{\.M}$ and low $v_{\infty}$. The spread in
excesses at any given $J-$band magnitude ($\sim0.2$ mag at $J-[3.6]$,
$\sim0.4$ mag at $J-[8.0]$) cannot be due to reddening, which would
displace all OB stars in a similar fashion. It is most likely related to
the range of mechanisms that produce free--free emission, the properties
of the winds ($\mbox{\.M}$, $v_{\infty}$, clumping factors) and stellar
rotation rates.

\vspace{-0.2cm}
Stars with $J>13$~mag that exhibit large excesses in
Figure~\ref{fig:oexcess} are classified as emission line Oe/On/One,
i.e. with evidence of fast rotation or circumstellar disks. In
Figure~\ref{fig:bexcess}, a larger fraction of stars with $J>13$~mag are
found to have large excesses, reflecting the higher fraction of emission
line stars among B-stars compared to O-stars \citep{Negueruela04}. The
``B extr'' stars \citep{Schild66, Garmany85, Conti86} are found to
occupy similar parts of the CMD as the Be(\ion{Fe}{2}) stars
\citep[notation by][]{Massey95} and Be-Fe stars \citep[notation
by][]{Evans06} (and also Be/X-ray binaries), suggesting these are
objects of one and the same nature, with the presence of \ion{Fe}{2}
lines perhaps due to higher densities in the circumstellar disks. We
note that some Be stars do not exhibit excesses, while several stars not
classified as Be stars have excesses. Variability (disk dissipation),
disk orientation, or insufficient spectral coverage leading to
inaccurate classifications can explain these outliers. Therefore,
infrared CMDs of optically blue selected stars can be used to identify
new Be stars. An alternate explanation for some of the outliers is the
new class of double periodic variables \citep[or DPVs;][]{Mennickent03,
Mennickent05, Mennickent08}. DPVs are evolved B and A-type giants in an
Algol mass-transfer configuration with circumprimary and circumbinary
disks. These disks are thought to be responsible for the observed long
period variability and infrared excess detected in the near-infrared.

\subsection{OB star SEDs}
\label{subsec:ob}

Ultimately, the position of a star in a two--color or color--magnitude
diagram is the result of its SED and luminosity, and it is the SED that
can help us interpret its position in terms of its physical
properties. Using the $0.3-24$~$\mu$m photometry from
Table~\ref{tab:phot}, we created SEDs for the stars by converting their
magnitudes to fluxes\footnote{Note, both magnitudes and fluxes exist in
the SAGE database, but only magnitudes are included in our photometric
catalog. Table 4 can be used to convert to fluxes.}. We used effective
wavelengths and calibrations from \citet{Bessell98} for $UBVI$,
\citet{Rieke08} for the 2MASS $JHK_s$ and IRSF, \citet{Reach05} for the
{\it Spitzer} IRAC bands, and the {\it MIPS Data
Handbook\footnote{http://ssc.spitzer.caltech.edu/mips/dh}} for the MIPS
24~$\mu$m band. We compared the observed SEDs with models to assess the
infrared excesses in the OB stars. TLUSTY models for LMC metallicity, a
micro-turbulent velocity of 10~km~s$^{-1}$, and the lowest available
surface gravity for each effective temperature $T_{\rm eff}$ were
selected to correspond to supergiant stars. The exact value of $\log g$
has little effect on mid-infrared fluxes for models with $T_{\rm eff}
\geq 25$kK and, while larger, the effect remains $\lesssim 0.1$~mag in
$(J-\lambda)$ (for $\lambda \geq 2.2\,\mu$m) for the cooler models.  We
did not attempt to fit the SEDs, but simply overplotted models to
provide a benchmark for determining the deviation of the observed SEDs
from a ``bare'' photosphere. Deviations of OB stars from a TLUSTY model
are due to reddening and free--free emission from their stellar winds.

Figure~\ref{fig:obstack} shows representative SEDs for 10 OB stars,
normalized by the flux at the $J-$band. The MCPS, IRSF and SAGE
measurements (filled circles), and the 2MASS and OGLE photometry (open
circles) are connected by a solid line. TLUSTY models, similarly
normalized, are overplotted as solid lines, while the dotted lines
correspond to the models reddened by $E(B-V) = 0.25$ and 0.50 mag (see
the Appendix for details on the treatment of interstellar
extinction). In Figure~\ref{fig:obresid}, we plot the same SEDs, divided
by the unreddened TLUSTY models. These unitless plots make deviations
from models obvious by avoiding the enormous flux ranges encountered
when plotting the SEDs of hot stars in the mid-infrared. This ease of
inspection comes with a price: when plotted against logarithmic
wavelength, equal bins on the abscissa do not correspond to equal energy
intervals. However, in these plots, the effects of reddening (clearly
seen in thet optical), infrared excess (detected in most cases) and even
variability between the 2MASS--IRSF and OGLE~III--MCPS photometry are
clearly seen.

The O stars shown in the left panels of Figures~\ref{fig:obstack} and
\ref{fig:obresid} include the lightly reddened, luminous and hot star Sk
$-70^\circ$~91, (O2~III(f*)+OB), which only has a weak infrared excess,
in spite of having very strong ultraviolet P-Cygni wind lines
\citep[$v_\infty =3150$~km~s$^{-1}$; e.g.][]{Massa03}. In such stars,
the wind density remains low, creating only weak infrared emission. In
contrast, the O4-5 V((f))pec star Sk~$-70^\circ$~60 and the O6~Iaf+ star
Sk~$-65^\circ$~22 (with $v_\infty = 1350$~km~s$^{-1}$) have strong
infrared excesses. The following two stars: Sk~$-66^\circ$~110 and
BI~192, are lightly reddened O9~I and O9~III stars, respectively, with
distinctive evidence for infrared emission from their winds. These
examples demonstrate that mass-loss rates cannot be derived solely from
infrared excesses; an estimate of the spectral type and luminosity class
is also required. We show examples of B giant and supergiant stars in
the right panels of Figures~\ref{fig:obstack} and \ref{fig:obresid}. The
B0 Ia star Sk~$-68^\circ$~52 is not consistent with the SED of a
reddened 30kK model, but shows evidence for infrared emission from a
wind. Contrast the B0~III star Sk $-67^\circ$~210, which appears to be
nearly unreddened, to the moderately reddened B0~III star ST2--46,
exhibiting infrared excess. Sk~$-68^\circ$~8 (B5~Ia+), compared to a
15kK model, similarly shows an infrared excess. The final SED is of
Sk~$-66^\circ$~12 (B8~Ia), also compared to a 15kK model (the coolest
TLUSTY model available).  While the star is probably cooler than 15kK,
the reduction of the effective temperature by a few kK cannot account
for the strong infrared excess, which is consistent with free--free
emission from a wind. Careful modeling of each star will be necessary to
accurately derive their wind properties.
 
\vspace{-0.4cm}
\section{Wolf-Rayet stars}
\label{sec:wr}
\vspace{-0.2cm}
Of the 125 WR stars in the catalog of \citet[][not including BAT99-106
and BAT99-108 through BAT99-112 in the crowded R136
cluster]{Breysacher99}, 99 yield matches in the SAGE
database\footnote{We label the B0 I + WN3 star Sk~$-69^\circ$~194
\citep{Massey00} as an early-B star in our plots.}. However, BAT99-45
and BAT99-83 are LBVs and BAT99-6 is an O star binary \citep[given its
O3f*+O spectral type;][]{Niemela01}. Only 3 of the remaining 96 WR stars
were detected at 24~$\mu$m: BAT99-22 (WN9h), BAT99-55 (WN11h), BAT99-133
(WN11h, a spectroscopic twin to the LBV BAT99-83 in minimum). Our
catalog includes various WR subtypes found in the LMC: early and late WN
stars (some with hydrogen present); early WC stars of WC4 subtype; and
one WO3 star. While most WR stars are thought to be post-main sequence
stars with progenitor masses greater than $25-75\,\msun$
\citep[depending on the subtype;][]{Crowther07}, there is increasing
evidence that the late-type WN stars with hydrogen in their spectra are
hydrogen-burning stars with initial masses above $40-60\,\msun$
\citep[the lower initial mass being uncertain;][]{Smith08}. Studies by
\citet{Bartzakos01}, \citet{Foellmi03}, and \citet{Schnurr08} found the
fraction of spectroscopic binaries among the WR stars in the LMC to be
lower than expected. Given the diverse properties of objects encompassed
under the ``WR'' classification, one might expect the position of WR
stars on a CMD or TCD to depend on their spectral types and possibly be
influenced by binarity.

We examine the position of the WR stars on the $J$ vs. $J-[3.6]$ CMD
(Figure~\ref{wrcmdjj36}) as a function of their spectral type,
subdividing them into: WN2-5, WN6-7, WN8, WN9-11, WC4 and WO3 and
labeling the known binaries. On average, the WN6-7 and WN9-11 stars are
more luminous and therefore brighter at $J$, when compared to their
early WN2-5 counterparts. In particular, the late WN9-11 stars
containing hydrogen are the brightest at $J$.  These are the most
luminous evolved WN stars, thought to be quiescent states of massive
LBVs \citep{Bohannan89, Smith94}. Overall, the WR stars span 6 mag in
$J$, and 2 mag in $J-[3.6]$ color, with the brighter stars being
``bluer'', i.e. with steeper SEDs (see below). The brightest star is
BAT99-22 (R84 or Sk $-69^\circ$~79) with a WN9h spectral type, which is
known to be blended with a red supergiant \citep{Heydari97}, while the
reddest star is BAT99-38 (HD 34602 or Sk $-67^\circ$~104), a triple
system \citep[WC4(+O?)+O8I:;][]{Moffat90} with an associated ring nebula
\citep{Dopita94}, which is probably responsible for the infrared
excess. Dust is known to form around some colliding wind binary WC stars
\citep[see references in][]{Crowther07}, and was recently detected in
the vicinity of WN stars \citep{Barniske08}. The 24~$\mu$m detections
around the WN11h stars BAT99-55 and BAT99-133, if confirmed, provide
additional examples of WN stars with associated dust.

In Figure~\ref{wrjkk80}, we examine the colors of the WR stars in a
$J-K_s$ vs. $K_s-[8.0]$ plot, following \citet{Hadfield07}. The location
of LMC WR stars does not agree with the region \citet{Hadfield07}
defined by dashed lines to represent colors of Galactic WR stars. There
are two reasons for this: 1) the Galactic sample includes many dusty WC9
stars that are intrinsically reddened \citep{Gehrz74b}, and 2)
line-of-sight reddening has scattered the Galactic sample into the
region these authors have defined. Metallicity is not expected to affect
the colors of WR stars, in particular WC stars, whose winds consist of
processed and therefore enriched material. The WR stars define a rather
narrow linear trend, independent of spectral type, defined by: $0 <
J-K_s < 0.7$ mag and $0.4< K_s-[8.0]<2.0$ mag, corresponding to power
law spectra with indexes $\alpha=1.5$ to 0.5, i.e. optically thick winds
with modest velocity gradients \citep[e.g.][]{Panagia91}. Single WR
stars are expected to have flatter SEDs than binaries, which typically
have main-sequence companions with steeper SEDs. The outliers in this
plot are once again the heavily reddened WC4 triple system BAT99-38 and
the blended star BAT99-22 (R84), whose infrared spectrum is dominated by
the red supergiant. R84 is an outlier in all the CMDs, located among the
RSG. On the $[3.6]-[4.5]$ vs. $[4.5]-[8.0]$ TCD (Figure~\ref{iraccc})
the WR stars follow a similar linear trend. We note that the brightest
WR stars in $J$ are the bluest in the TCDs.

In Figure~\ref{fig:wr-lbvstack}, we show SEDs for 7 WR stars with
representative subtypes. The strong \ion{C}{4} and \ion{He}{2} emission
lines present in the WO3 and WC4 stars \citep[see][]{Breysacher99} are
responsible for the ``kinks'' in the SEDs of these stars. For 3 of these
stars (BAT99-123, BAT99-61, BAT99-17) we show model fits derived from
mainly optical and near-infrared spectroscopy using the spherical,
non-LTE, line-blanketed CMFGEN code of \citet{Hillier98}. For BAT99-123
(WO3), we use the model presented in \citet{Crowther00} which is derived
from line profile fits to ultraviolet and optical spectroscopy. This
model has parameters of $T_{\rm eff} = 150$ kK, log $L/L_\odot = 5.3$ ,
$v_\infty = 4,100$ km\,s$^{-1}$ and $\dot M = 1 \times 10^{-5}\,\msun$
yr$^{-1}$ for a volume filling factor of 0.1. For BAT99-61 (WC4), we use
the model presented by \citet{Crowther02} which has $T_{\rm eff} = 52$
kK, log $L/L_\odot = 5.68$, $v_\infty = 3,200$ km\,s$^{-1}$ and $\dot M
= 4 \times 10^{-5}\, \msun$ yr$^{-1}$ for a volume filling factor of
0.1.  For BAT99-17 (WN4o), we use the model summarized in
\citet{Crowther06b} which has $T_{\rm eff} = 52$ kK, log $L/L_\odot =
5.40$ , $v_\infty = 1,750$ km\,s$^{-1}$ and $\dot M = 1.4 \times
10^{-5}\, \msun$ yr$^{-1}$ for a volume filling factor of 0.1.  Overall,
the model fits presented in Figure~\ref{fig:wr-lbvstack} show that the
observed SEDs in the mid-infrared are flatter than the model continua,
suggesting that either the model mass loss rates are too low or the
winds are more highly clumped than the models predict. We caution,
however, that this comparison is only for 3 stars. Generic CMFGEN models
\citep{Smith02} are shown for the other 4 WR stars: a WN 85kK model for
BAT99-1 (WN3b) and a WN 45kK model for the remaining WNL stars, all of
which show excess above that predicted by the models. It is clearly
desirable to investigate these apparent differences further by
performing detailed line profile fits in the mid-infrared.

Finally, in Table~\ref{tab:mips70} we present photometry for 10 WR stars
with counterparts (within 20$\arcsec$) at 70~$\mu$m and for 1 WR star
with a counterpart at 160~$\mu$m. The star name, spectral type,
magnitude and associated error at 70~$\mu$m and 160~$\mu$m, followed by
the flux and associated error in these bands, are given. We caution that
the angular resolution of MIPS is $18\arcsec$ and $40\arcsec$,
respectively, at these bands. However, we note that the sgB[e] and LBVs
detected also show evidence for dust at the IRAC+MIPS24 bands.

\section{Luminous Blue Variables}
\label{sec:lbv}

\begin{figure}[ht]  
\includegraphics[width=6in]{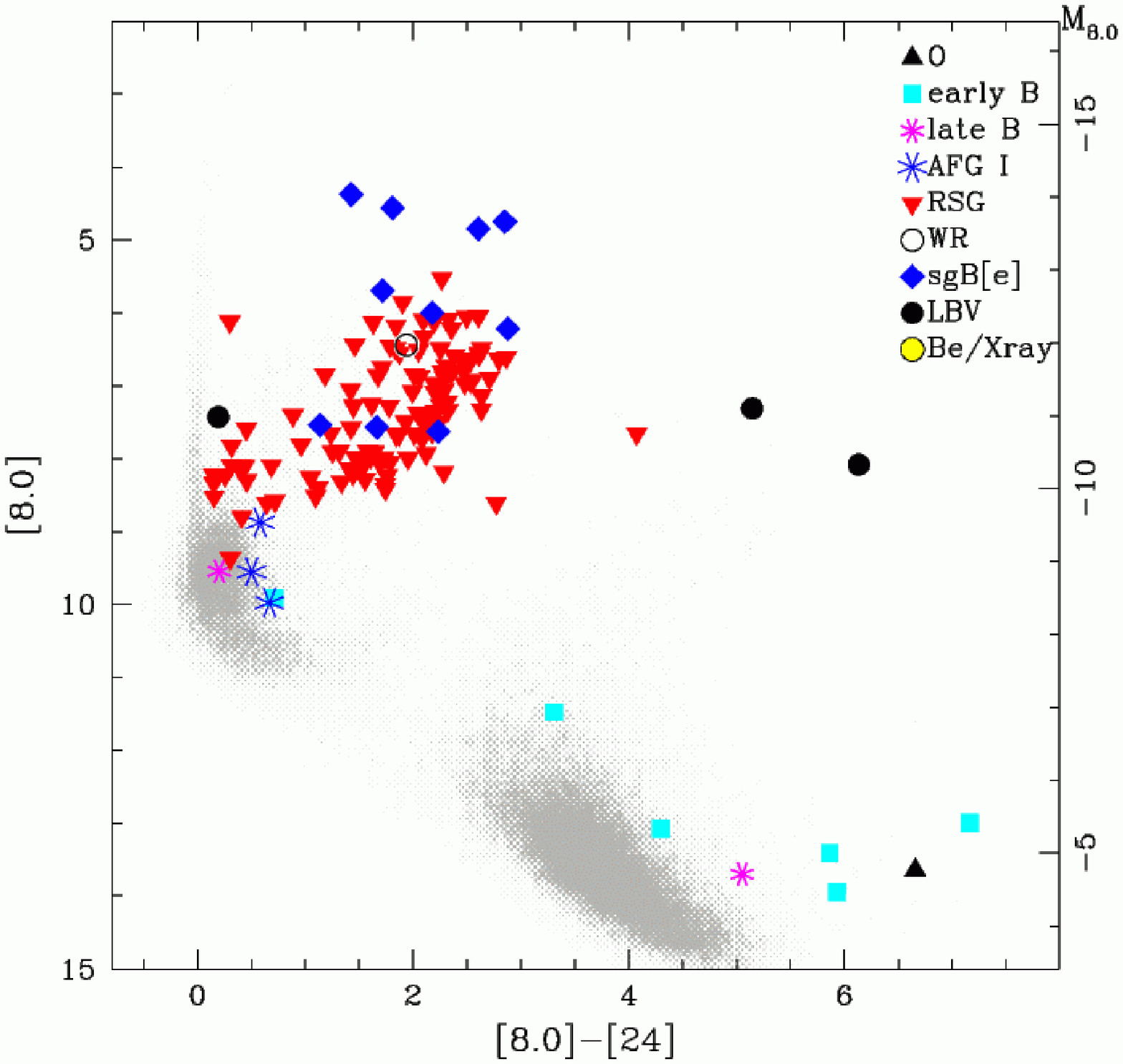}
\caption{Same as Figure~\ref{cmd36}, but for the [8.0] vs. $[8.0]-[24]$
color magnitude diagram. The sgB[e], RSG and LBVs are also among the
most luminous stars at 8~$\mu$m.}
\label{cmd8}
\end{figure}

\begin{figure}[ht]  
\includegraphics[width=6in]{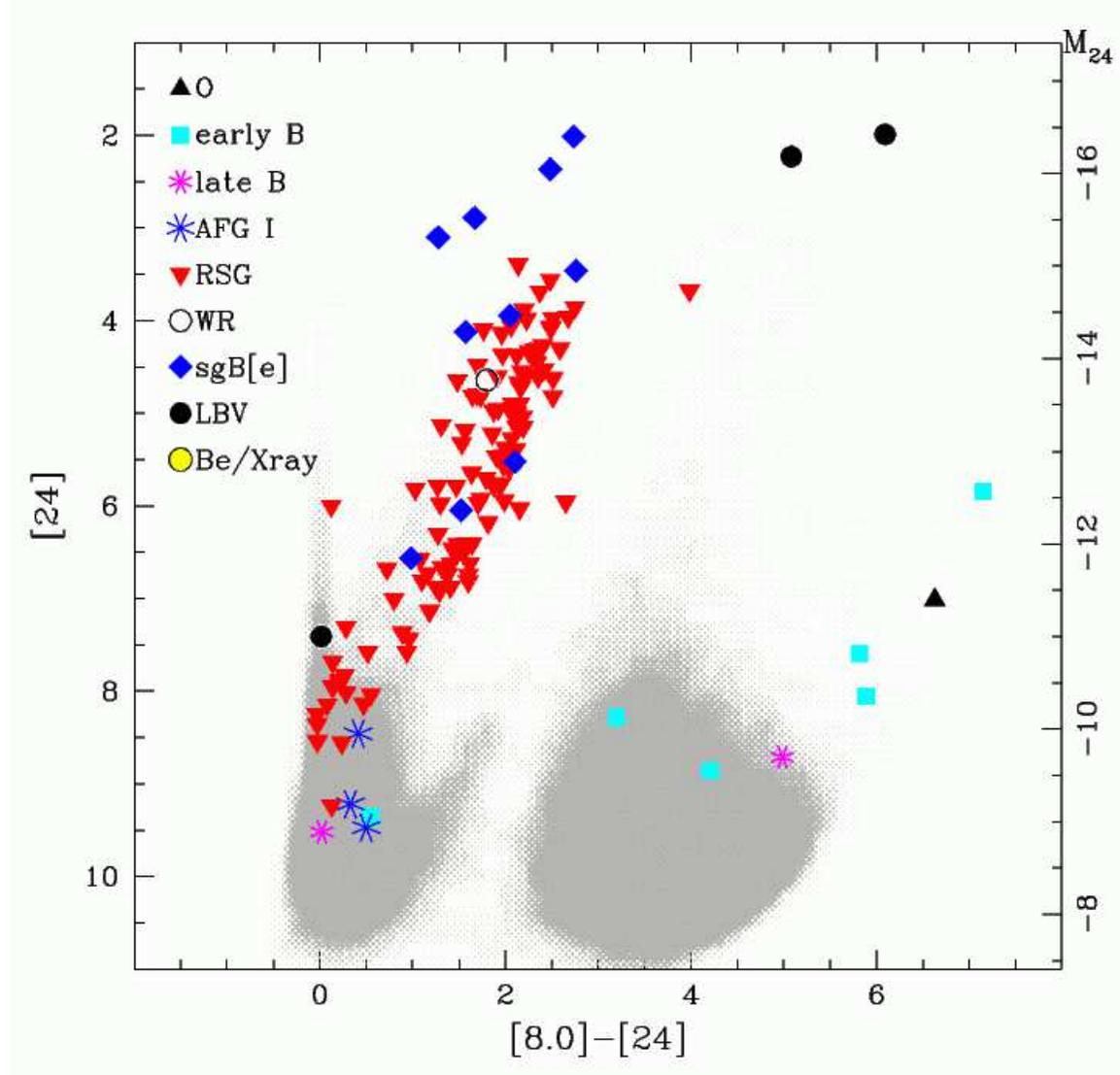}
\caption{Same as Figure~\ref{cmd36}, but for the [24] vs. $[8.0]-[24]$
color magnitude diagram. The brightest sgB[e], RSG and LBVs are among
the most luminous stars at 24~$\mu$m.}
\label{cmd24}
\end{figure}

There are 6 confirmed LBVs \citep[see review by][]{Humphreys94} in the
LMC: S~Dor, BAT99-83 or R127, R~71, R~110, BAT99-45, and R~85, although
another 5--6 have been suggested as candidates in the literature
\citep[see e.g.,][]{Weis03}. The LBVs are not only among the most
luminous sources at 3.6~$\mu$m (see Figure~\ref{cmd36} and
\ref{cmd36j36}), with [3.6]--[4.5] colors similar to AGB stars and
intermediate between RSG and sgB[e] stars, but also at 8.0~$\mu$m
(Figure~\ref{cmd8}) and 24~$\mu$m (Figure~\ref{cmd24}).  In the TCDs
(Figures~\ref{jkk80} and \ref{iraccc}) the LBVs are located between the
OB and WR stars, with the exception of R71 \citep{Lennon93}, which is an
extreme outlier, with $K_s-[8.0]=3.4$~mag and $[4.5]-[8.0]=2.5$~mag. Its
recent brightening of 1.5 mag between 2006 and 2009 \citep[see light
curve from the {\it All Sky Automated
Survey}\footnote{http://www.astrouw.edu.pl/asas/};][]{Pojmanski02}
cannot account for its infrared colors. Instead, the emission from
polycyclic aromatic hydrocarbons (PAHs) detected by \citet{Voors99} in
R71 can explain its ``red'' colors, since there are strong PAH lines in
the [3.6] and [8.0] bands. All 6 confirmed LBVs were detected in the
IRAC bands, 3 were detected in the MIPS 24~$\mu$m band and 2 at
70~$\mu$m: R71 (saturated at 24~$\mu$m) and BAT99-83 or R127 (see
Table~\ref{tab:mips70}). The R71 detection is consistent with the
60~$\mu$m value in the IRAS Point Source Catalog and \citet{Voors99}
suggest that a combination of crystalline silicates and cool dust are
responsible for it. The detection of R127 at 70~$\mu$m may imply that a
similar dusty environment exists around this star. We note that BAT99-45
is distinctively ``bluer'' than the other two LBVs detected at
8.0~$\mu$m and 24~$\mu$m (see Figures~\ref{cmd8}, \ref{cmd24},
\ref{iraccc2}).

The SEDs of all 6 confirmed LBVs are shown in
Figure~\ref{fig:wr-lbvstack} and, unlike other classes of stars, are
highly non-uniform. Since LBVs are by definition variable and the
photometry in different bands was obtained at different epochs, the
differences in the 2MASS and IRSF photometry are most likely due to
variability. We only have MCPS photometry for two LBVs, which seems
either variable or of suspect quality. S~Dor, the prototype S Doradus
variable, and R127 \citep[in outburst during the SAGE
observations;][]{Walborn08} are the brightest LBVs at 24~$\mu$m
(Figure~\ref{cmd24}). Many LBVs are surrounded by small nebulae (0.2--2
pc) originating from past giant eruptions, which form cool dust. For
example, R127 was found to have a bipolar nebula \citep{Weis03}. The
dust associated with its nebula is likely responsible for the bright
mid-infrared magnitudes of R127. R71 exhibits a strong excess in the
IRAC bands, as expected from its position in the TCDs.

Do LBVs represent a homogeneous class of objects? Their infrared SEDs
point to diverse properties, which is not surprising given their
spectral variability during eruptions. On the one hand, the low
luminosity LBV R71, with an estimated progenitor mass of around
40~$\msun$ has one of the strongest infrared excesses of any massive
star in the LMC and is well known as a source of dust emission. S~Dor
and R127 are more typical massive LBVs, both of which exhibit an
infrared excess, possibly linked to circumstellar material formed during
their recent outbursts. The other three LBVs in the sample: R110,
BAT99-45 (also Sk~$-69^\circ$~142a, S83) and R85 all show different but
coherent infrared behaviour more consistent with a stellar wind
contribution only. It is possible that dust around these objects is
cooler; however these stars were not detected at the longer MIPS
bands. The different shapes of the LBV SEDs are likely related to the
time since the last outburst event and the amount of dust formed.

\section{B[e] supergiants and Classical Be stars}
\label{sec:sgBe}

B[e] supergiants (sgB[e]) are a distinct class of B-type stars
exhibiting forbidden emission lines, or the ``B[e] phenomenon''
\citep{Lamers98}. They are characterized by strong Balmer emission
lines, narrow permitted and forbidden low-excitation emission lines and
a strong mid-infrared excess. A two-component wind model with a hot
polar wind and a slow and cool equatorial disk wind has been proposed to
explain them \citep{Zickgraf86}; however models of the disk have so far
achieved limited success \citep{Porter03, Kraus07, Zsargo08}. On the
Hertzsprung-Russell (HR) diagram, sgB[e] stars are located below the
Humphreys-Davidson limit \citep{Humphreys79b}, a few of them being
coincident with the location of LBVs. However, the existence of an
evolutionary connection of sgB[e] stars to LBV stars (transitional phase
between Of and WR stars) is unclear. \citet{Langer98} outline three
possible ways sgB[e] could form circumstellar disks: from single massive
evolved stars close to critical rotation \citep{Meynet06}, blue
supergiants which have left the red supergiant branch, and from massive
binary mergers \citep{Podsiadlowski06}.

The sgB[e] stars are the most conspicuous group of stars in all infrared
CMDs and TCDs: they are among the brightest and most reddened stars in
the LMC \citep[also noted by][]{Buchanan06}. In the LMC, 12 stars have
been classified as sgB[e] stars \citep[including LH 85--10, although it
is not among the 11 stars listed in][]{Zickgraf06}. The 11 that were
included in our catalog (S~22, S~134, R~126, R~66, R~82, S~12, LH~85-10,
S~35, S~59, S~137, S~93) were all matched in the SAGE database. Their
SEDs are shown in Figure~\ref{fig:sgbestack}. The B0.5[e] stars are
sorted by decreasing flux at [3.6], the later-type B[e] stars by
decreasing effective temperature. A 25kK TLUSTY model representing the
underlying photosphere is overplotted. We find that the SED of LH
85--10, a suspected LBV assigned a B[e] spectral type by
\citet{Massey00}, resembles the SEDs of ``classical Be'' stars
\citep{Negueruela04b}. Figures~\ref{cmd36}, \ref{cmd36j36}, \ref{jkk80},
and \ref{iraccc} demonstrate that LH 85--10 is much fainter and less
reddened than sgB[e] and LBV stars, and is located near Be stars, and
therefore may have been misclassified as a B[e] star. The SEDs of the
remaining 10 sgB[e] stars are all very similar, with slowly decreasing
flux in the optical, an inflexion point in the near-infrared and a
``bump'' starting at 2~$\mu$m and peaking near 5~$\mu$m. This peak
corresponds to hot dust at $\sim$600~K. The slight change in the slopes
of the SEDs between 8 and 24~$\mu$m from star to star suggests different
contributions from cool dust (150~K). R66, S35 and R126 were also
detected at 70~$\mu$m (Table~\ref{tab:mips70}). We find all 10 sgB[e]
stars to be extincted by $E(B-V)\gtrsim0.5$.

\citet{Kastner06} presented {\it Spitzer} IRS spectroscopy for 2 of
these sgB[e] (R66, R126\footnote{Note, the sgB[e] star R126 (Sk
$-69^\circ$ 216) is located in NGC 2050, close to the LBV R127 (Sk
$-69^\circ$ 220), which is in NGC 2055. These two clusters could be a
``fossil'' two-stage starburst.}), finding evidence for massive
circumstellar disks, which can simultaneously explain the (relatively)
small amount of reddening in the optical and the observed infrared
excess. However, the mass-loss rate required to sustain such a massive
disk is very high and unlikely to persist during the He burning phase of
the star. These authors argued against a model of episodic mass ejection
based on the similarity of the SEDs of the two sgB[e] they studied. The
similarity of all 10 sgB[e] SEDs makes the scenario of episodic mass
ejection even less likely or, alternatively, suggests a shorter duration
of the sgB[e] phase or even an origin in a binary companion. The small
fraction of sgB[e] stars among B stars might present a clue to their
origin. We note that the rare, dust obscured, luminous progenitors
\citep{Prieto08, Prieto08b} of the newly discovered class of transients,
including SN2008S \citep{Smith09} and NGC~300~OT \citep{Bond09,
Berger09}, could be related to the lower luminosity B[e] supergiants,
given that both are likely post-main-sequence stars (possibly in a
post-red supergiant phase) with dust and similar
luminosities. \citet{Thompson08} presented evidence for a short duration
($< 10^4$ yr) of the dust-enshrouded phase prior to eruption.

The infrared emission in the sgB[e] stars is much stronger than in
classical Be stars (shown in Figure~\ref{fig:bestack}). It is typically
several hundred times stronger than the photosphere at 8~$\mu$m, and up
to 1000 times stronger at 24~$\mu$m, explaining the extreme positions of
sgB[e] stars in the TCDs. In the classical Be stars, free--free emission
from disks can cause the 8 $\mu$m\ flux to be 10 times that of the
underlying photosphere, as is the case for all but one
(Sk~$-71^\circ$~13) of the Be stars shown. The emission in classical Be
stars can be highly variable \citep[see review by][]{Porter03a}, which
may account for the disagreement between the IRSF and 2MASS photometry
for NGC~2004-033 (B1.5e), and in some cases, can even dissipate
altogether, as for Sk~$-71^\circ$~13 (Be~(Fe-Be)). The SEDs of the
remaining Be, ``B~extr'', Be (Fe-Be) stars are identical, thereby
providing additional evidence (see \S\ref{sec:ob}) that these stars are
the same objects: the most luminous of the Be stars. We also present the
SED of 1 of the 4 Be/X-ray binaries in our matched catalog, finding it
to have a peculiar shape, similar to the other Be stars, but with excess
emission in the [3.6] and [4.5] bands, possibly due to variability.

\section{Red Supergiants \& Other Stars}
\label{sec:other}

\begin{figure}[ht]  
\plotone{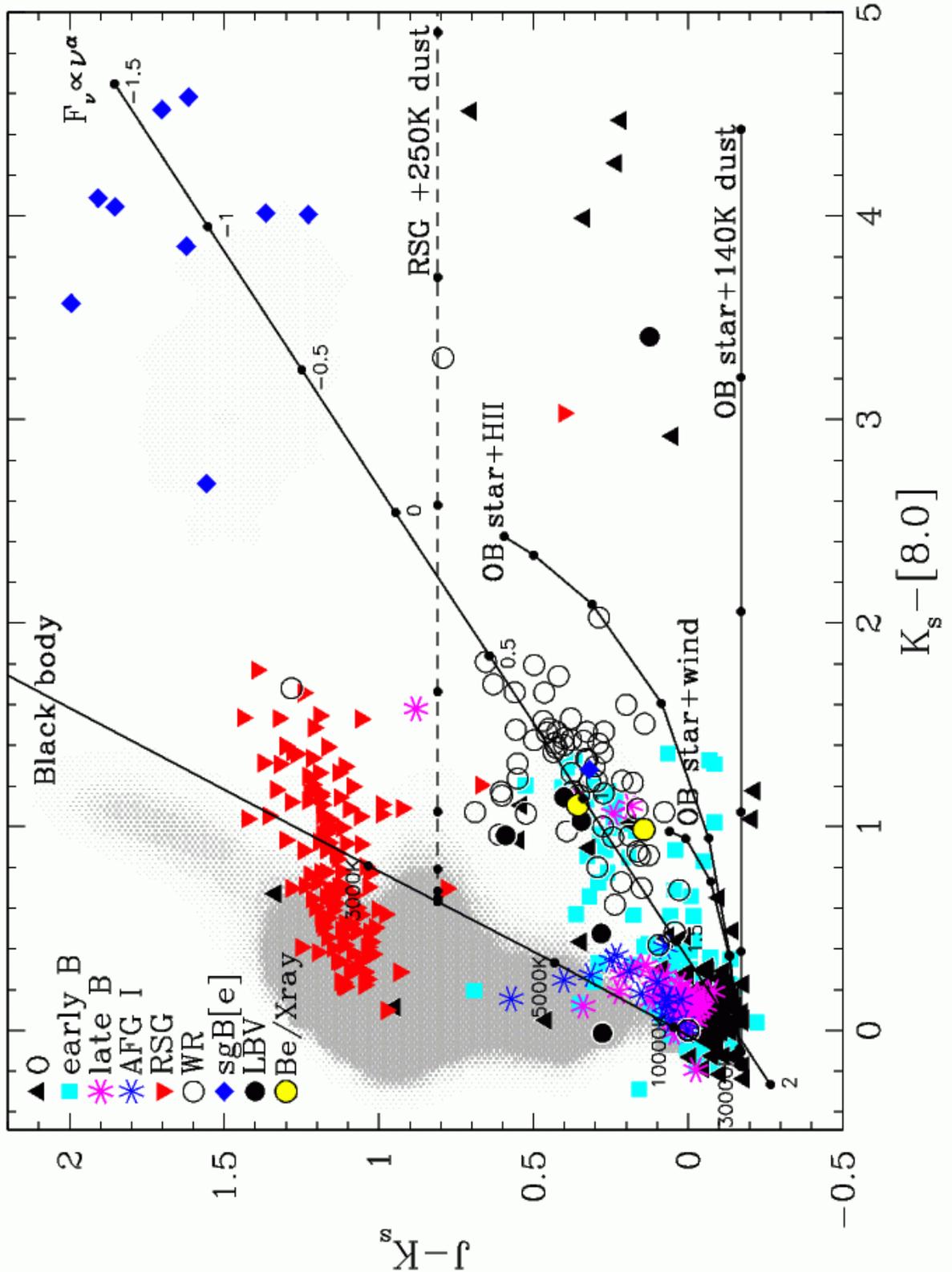}
\caption{$J-K_s$ vs. $K_s-[8.0]$ diagram for massive stars in our
catalog. The locations of all the SAGE detections are shown in grey as a
Hess diagram. The solid lines represent models (described in
\S\ref{sec:colorcolor}): (i) a BB at various temperatures, as labelled,
(ii) a power law model F$_\nu \propto \nu^{\alpha}$, for
$-1.5\leq\alpha\leq2$, (iii) an OB star plus an ionized wind (not
labelled), (iv) an OB star plus emission from an optically thin HII
region, (v) an OB star plus 140~K dust, (vi) 3,500~K blackbody plus
250~K dust (dashed line). The sgB[e], RSG and WR stars occupy distinct
regions on this diagram.}
\label{jkk80}
\end{figure}

\begin{figure}[ht]  
\plotone{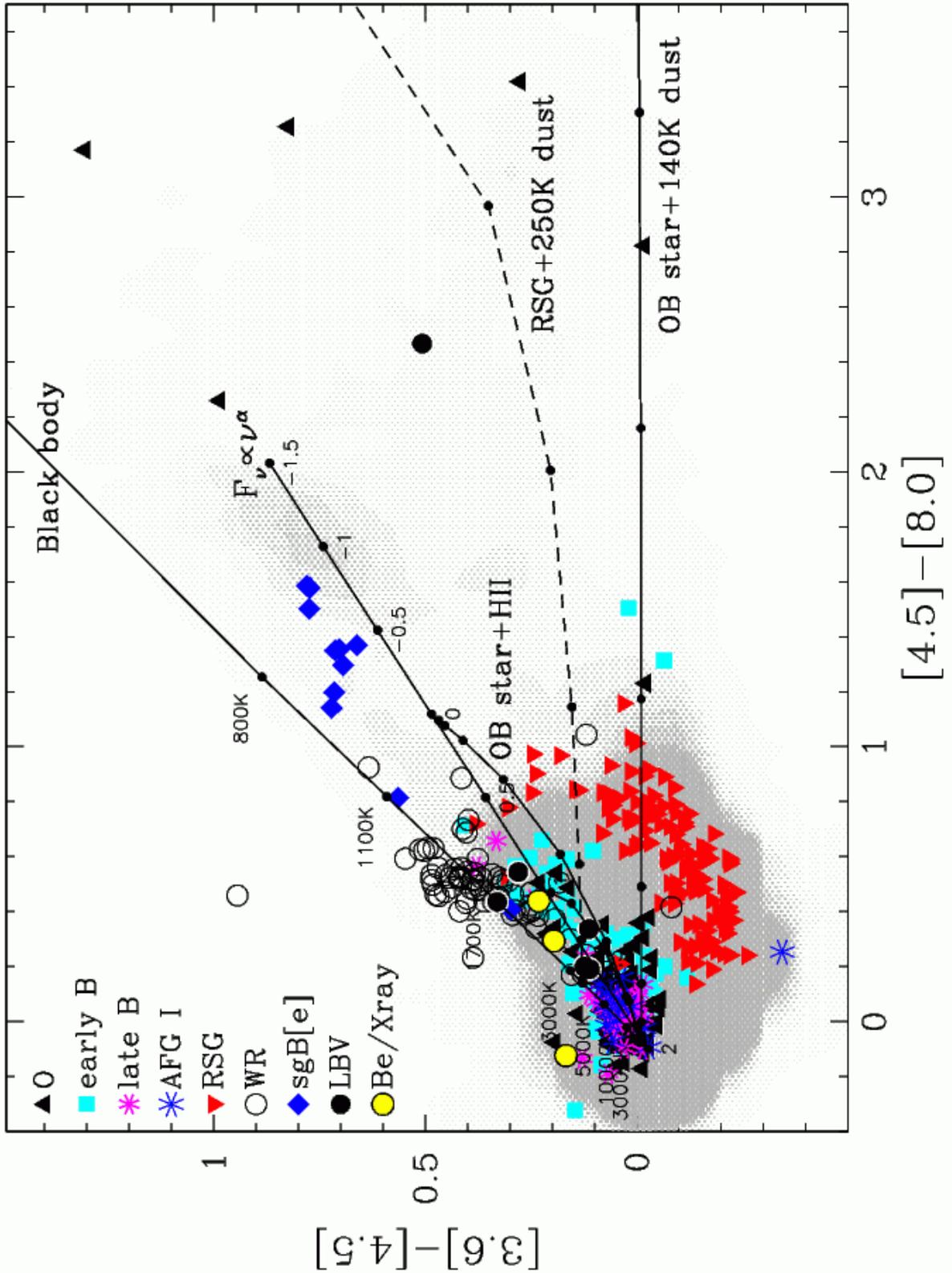}
\caption{Same as Figure~\ref{jkk80}, but for the $[3.6]-[4.5]$
vs. $[4.5]-[8.0]$ diagram. The majority of hot massive stars lie between
the blackbody and OB star +wind model, illustrating that a BB is a good
approximation in the infrared.}
\label{iraccc}
\end{figure}

\begin{figure}[ht]  
\plotone{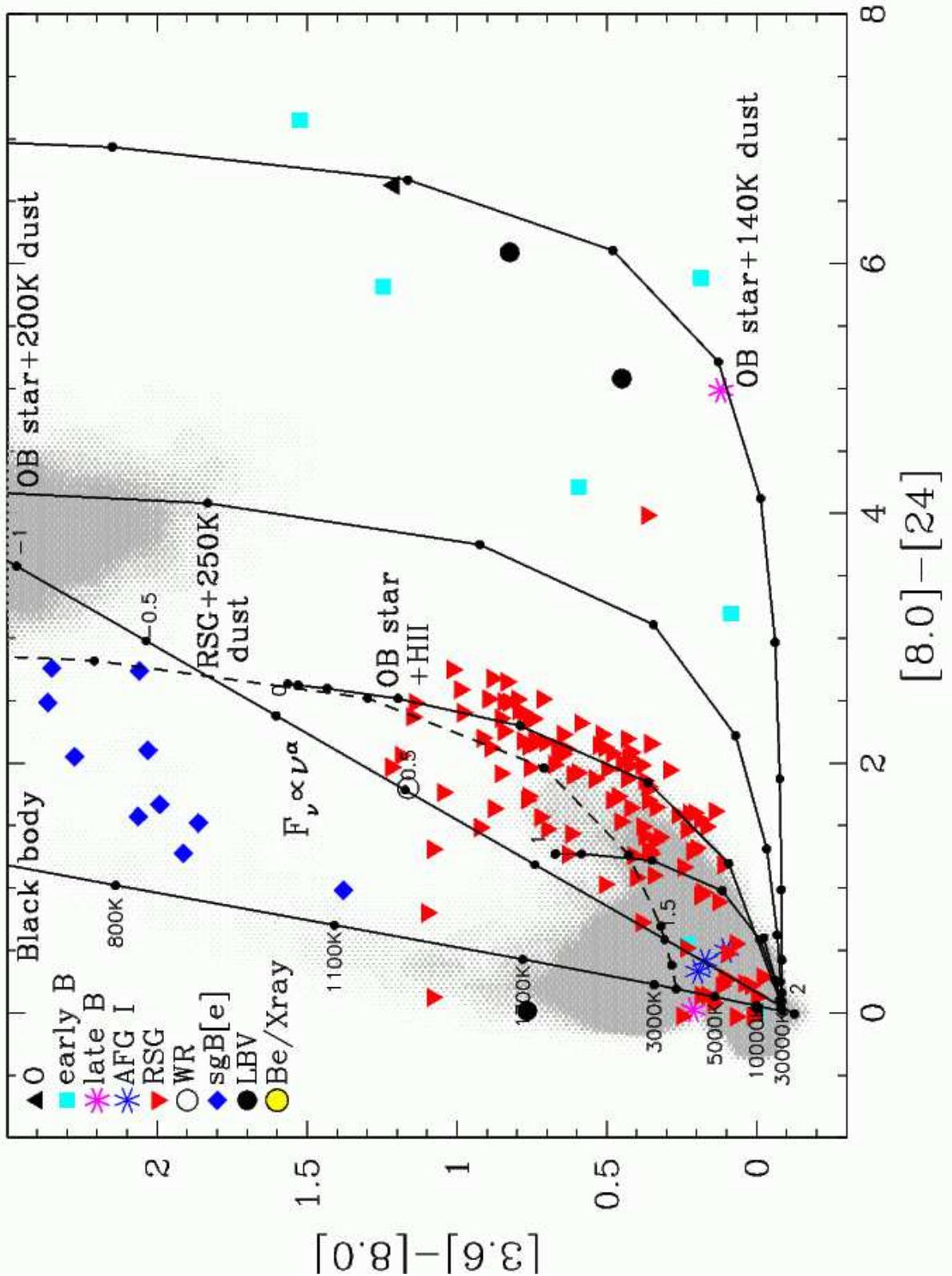}
\caption{Same as Figure~\ref{jkk80}, but for the $[3.6]-[8.0]$
vs. $[8.0]-[24]$ diagram. The RSG with ``bluer'' colors have earlier
spectral types, indicating a temperature sequence (see
\S\ref{sec:other}).}
\label{iraccc2}
\end{figure}

\begin{figure}[ht]  
\includegraphics[width=6in]{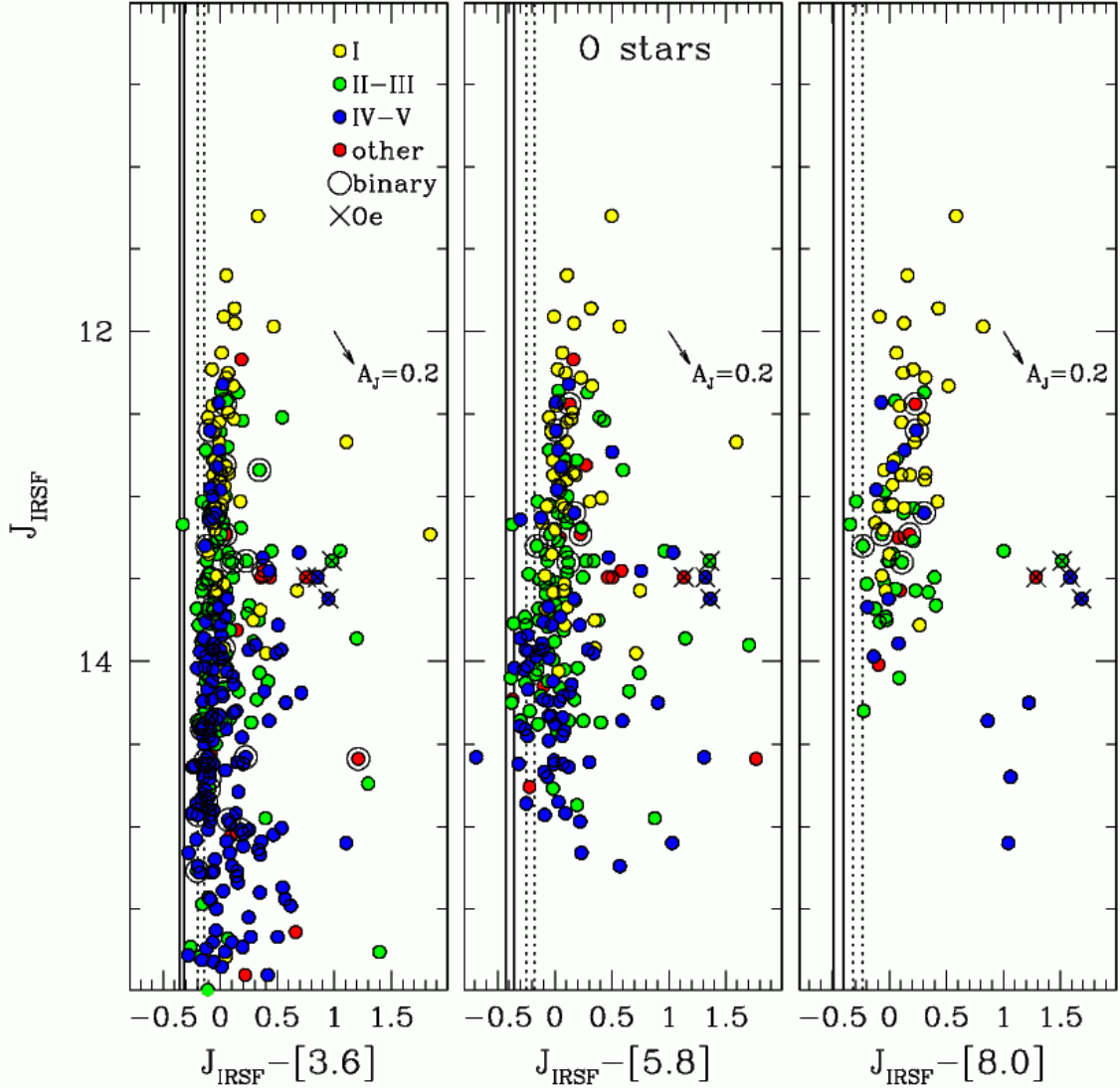}
\caption{Infrared excesses ($J_{IRSF}$ vs. $J_{IRSF}-[3.6]$,
$J_{IRSF}-[5.8]$ and $J_{IRSF}-[8.0]$) for 354 O stars. Supergiants are
shown in yellow, giants in green, main-sequence stars in blue, stars
with uncertain classifications (``other'') in red, binaries with a large
circle and Oe stars with an $\times$. The solid lines correspond to 30kK
and 50kK TLUSTY models with $\log g = 4.0$. A reddening vector for
$E(B-V)=0.2$ mag is shown, as well as reddened TLUSTY models by this
same amount (dotted lines). The more luminous stars exhibit larger
infrared excesses, which increase with $\lambda$. The $0.2-0.4$ mag
spread in excesses at any $J_{IRSF}-$band magnitude reflects the range
in mass-loss rates, terminal velocities, clumping properties and,
perhaps, rotation rates of O stars.}
\label{fig:oexcess}
\end{figure}

\begin{figure}[ht]  
\includegraphics[width=6in]{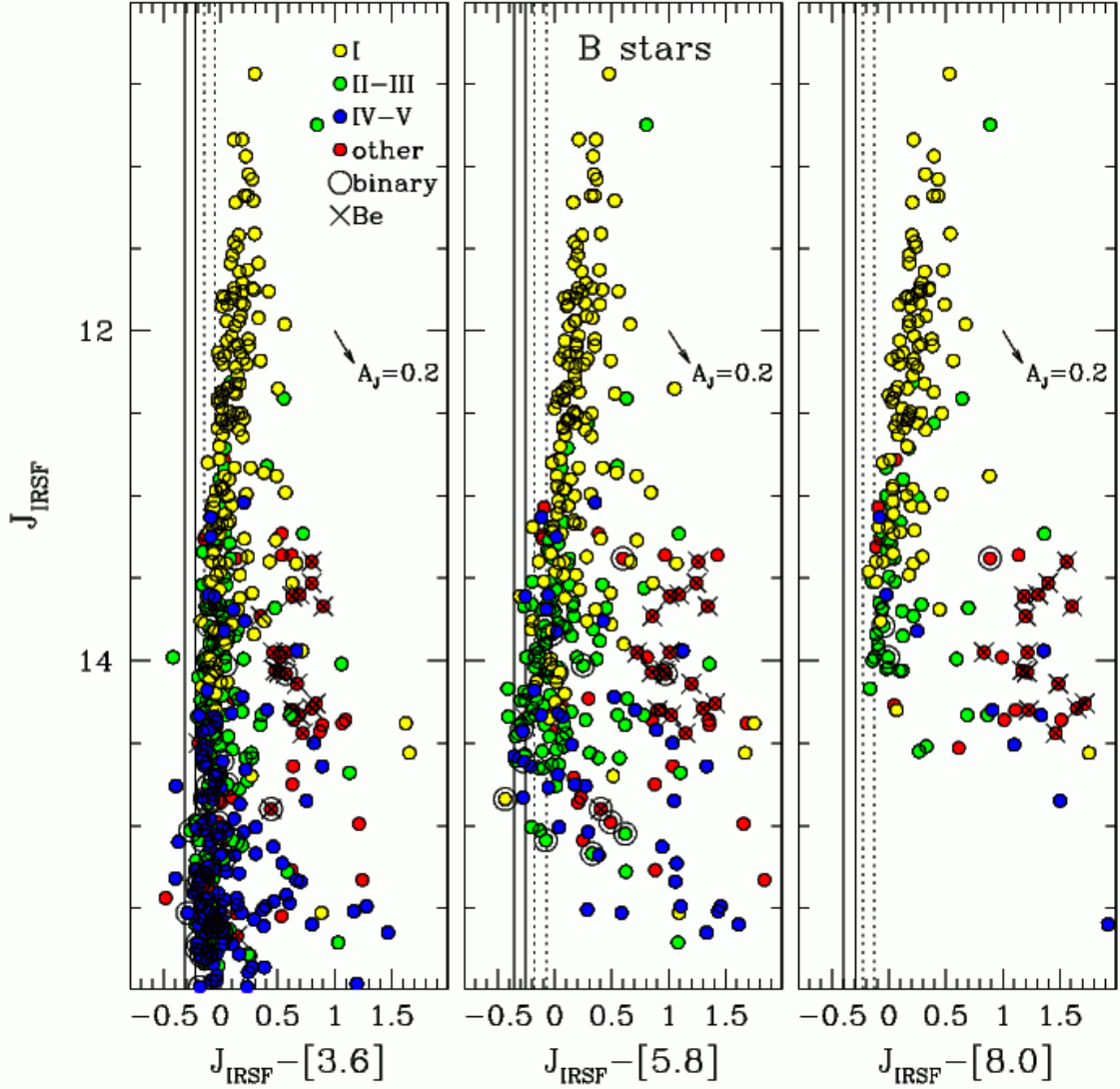}
\caption{Same as Figure~\ref{fig:oexcess}, but for 586 early-B
stars. The solid lines correspond to 20kK, $\log g = 3.0$ and 30kK,
$\log g = 4.0$ TLUSTY models. A reddening vector for $E(B-V)=0.2$ mag is
shown, as well as reddened TLUSTY models by this same amount (dotted
lines). The larger number of early-B stars makes the trends identified
among the O stars clearer.}
\label{fig:bexcess}
\end{figure}

\begin{figure}[ht]
\epsscale{1.00}
\plottwo{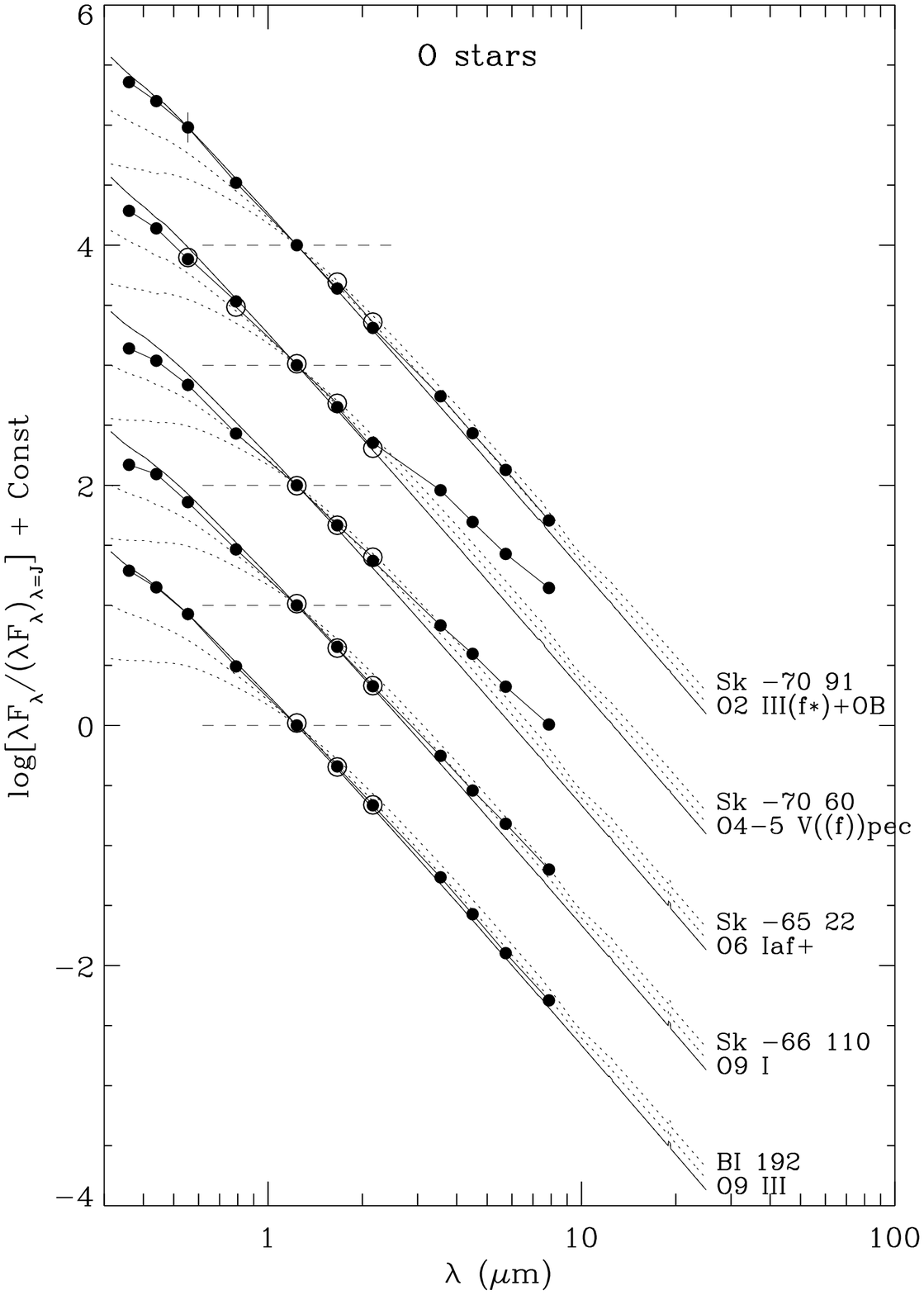}{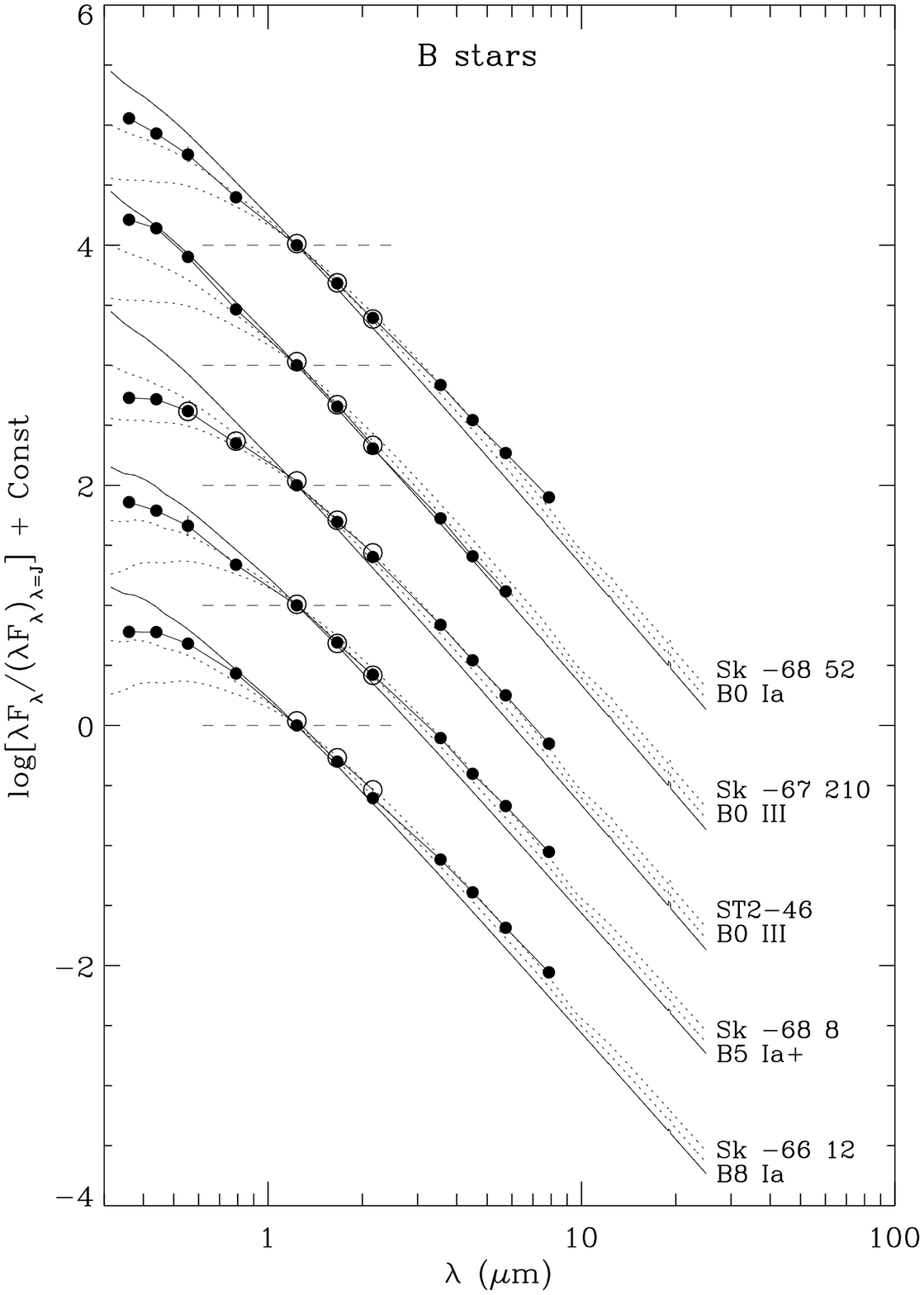}
\caption{Representative SEDs of O stars $(Left)$ and B stars $(Right)$,
normalized by their $J-$band fluxes (dashed line) and offset for display
purposes. Normalized TLUSTY model atmospheres (30kK, $\log g = 3.0$
model for the O and B0 stars; 15kK, $\log g = 1.75$ model for the late B
stars) are overplotted for comparison. The MCPS, IRSF and SAGE
measurements are shown as filled circles; the 2MASS and OGLE
measurements as open circles. The dotted curves correspond to TLUSTY
models reddened by $E(B-V) = 0.25$ and 0.50 mag. Infrared excesses are
detected in most stars (see \S\ref{subsec:ob}).
\label{fig:obstack}}
\end{figure}

\begin{figure}[ht]
\epsscale{1.00}
\plottwo{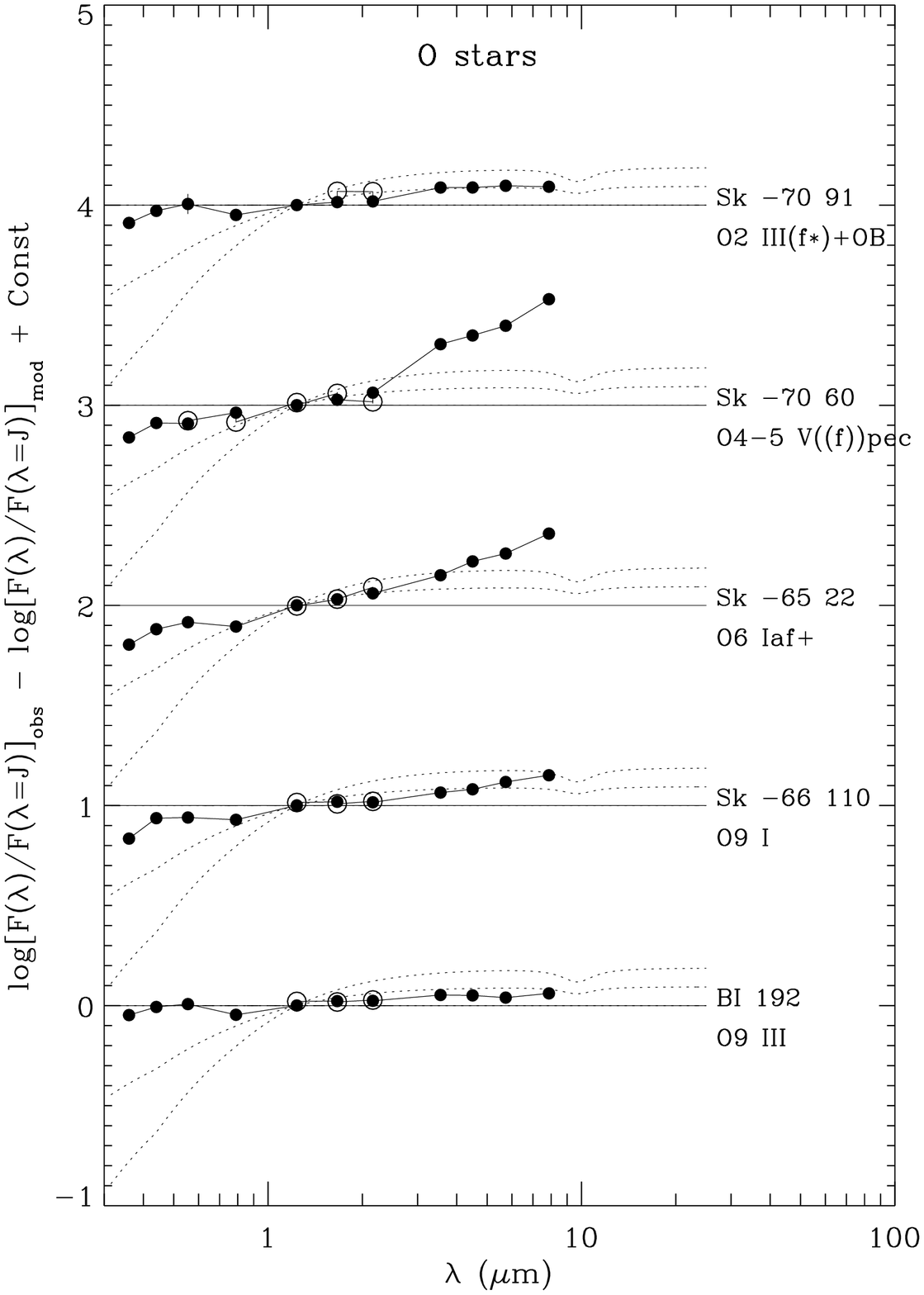}{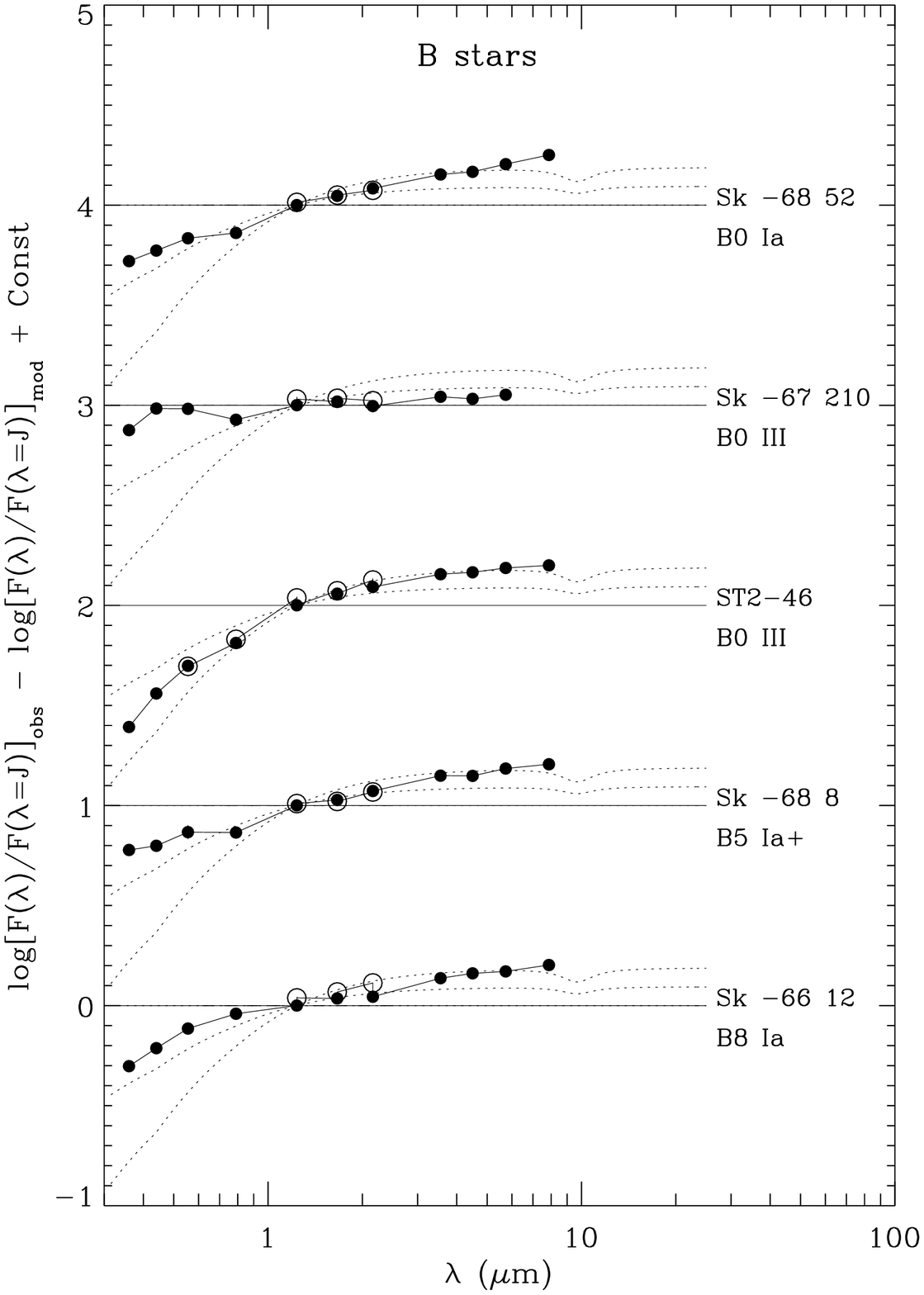}
\caption{Ratios of the SEDs of O stars $(Left)$ and B stars $(Right)$ to
the unreddened TLUSTY models shown in Figure~\ref{fig:obstack}, clearly
showing the deviations of the observations from the models and cases of
variability, from a comparison of $VIJHK_s$ magnitudes obtained from
different sources. Dotted curves correspond to TLUSTY models reddened by
$E(B-V) = 0.25$ and 0.50 mag. Infrared excesses are present in most
cases.
\label{fig:obresid}}
\end{figure}

\begin{figure}[ht]  
\includegraphics[width=6in]{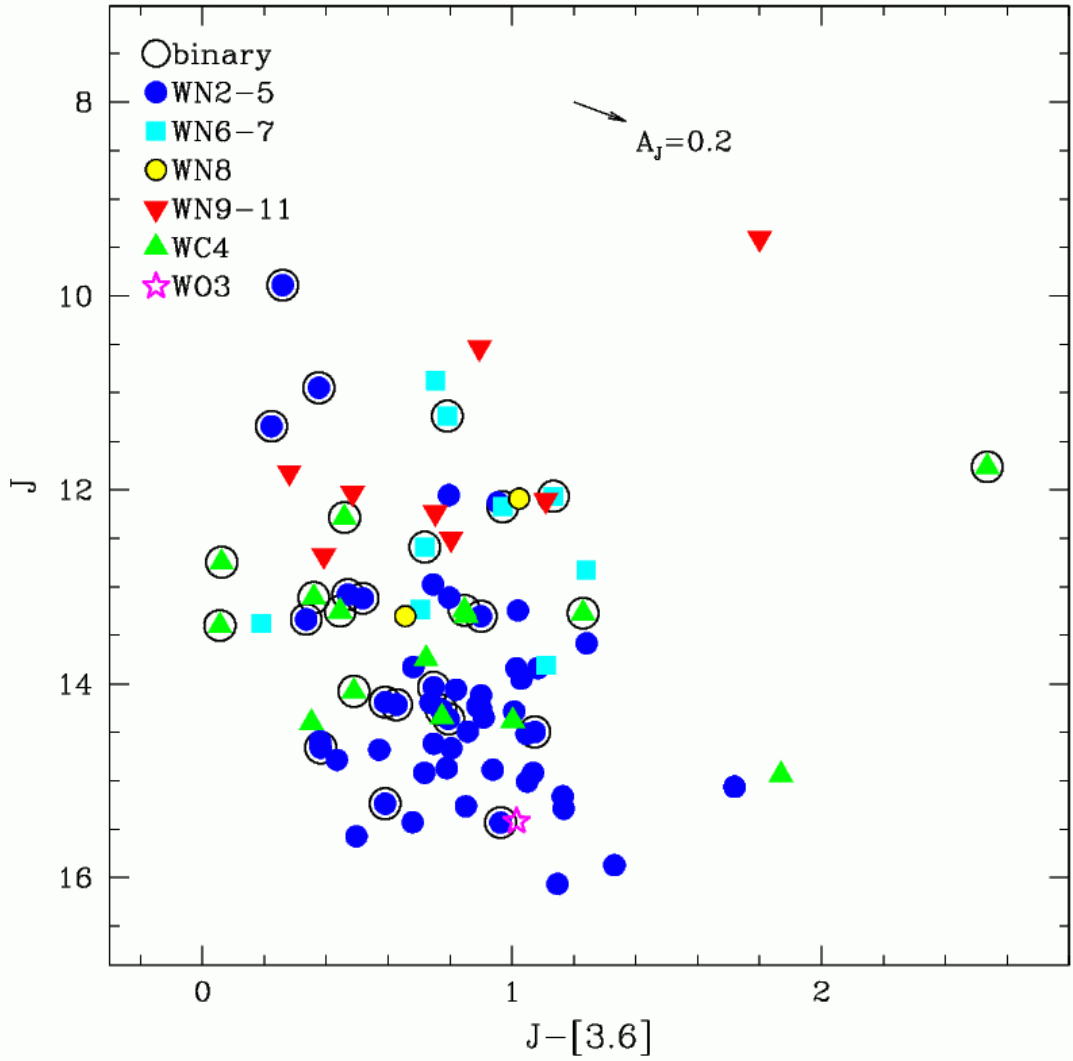}
\caption{$J$ vs. $J-[3.6]$ color magnitude diagram for Wolf-Rayet stars
with infrared detections, labelled according to spectral type. Black
circles denote known binaries. The reddening vector for $E(B-V)=0.2$ mag
is shown. The WN2-5 stars are on average fainter, while the WN6-7 and
WN9-11 are brighter at $J-$band.}
\label{wrcmdjj36}
\end{figure}

\begin{figure}[ht]  
\includegraphics[width=6in]{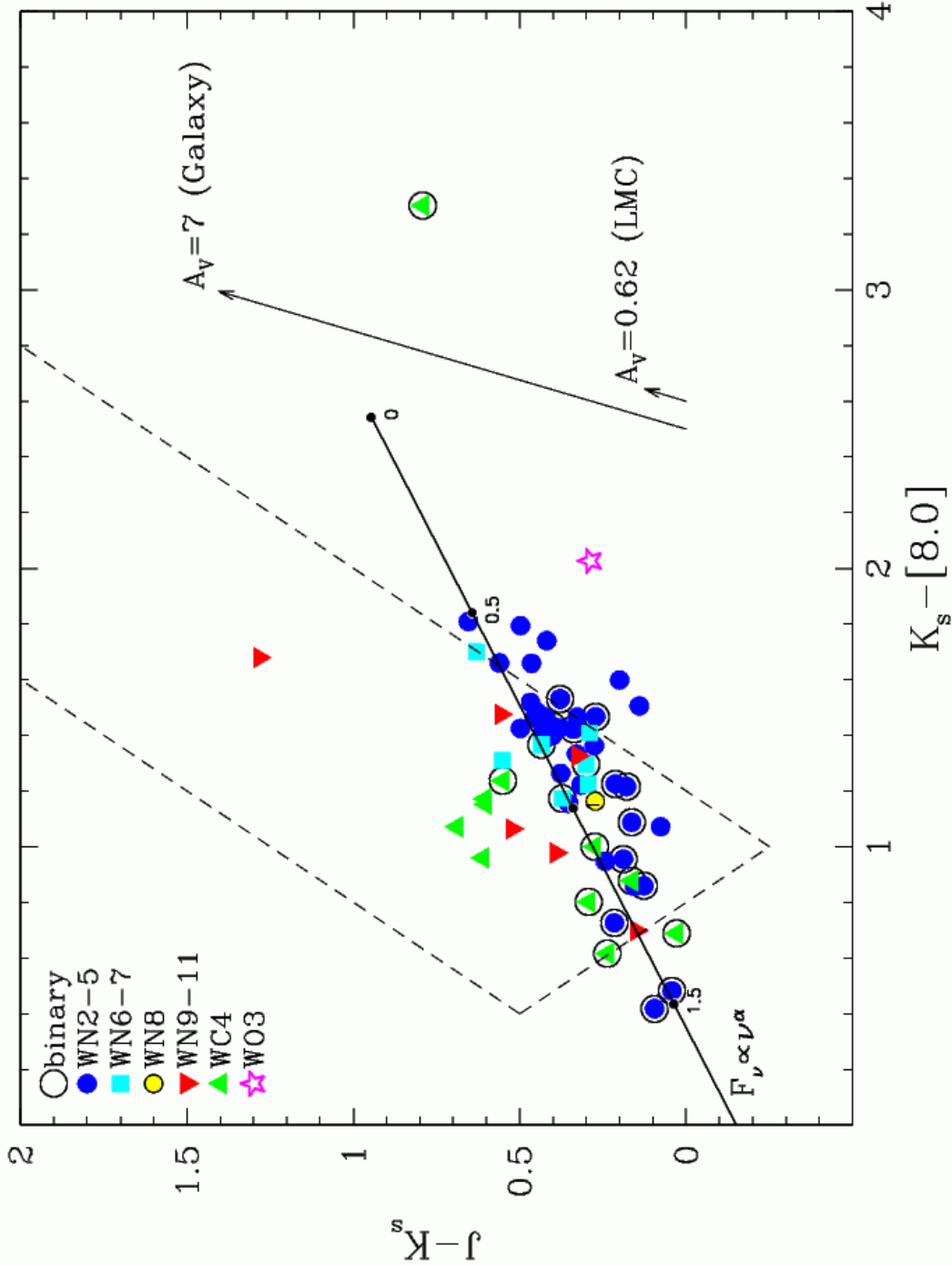}
\caption{$J-K_s$ vs. $K_s-[8.0]$ diagram for Wolf-Rayet stars with
infrared detections, showing a tight linear correlation independent of
spectral type. The dashed lines correspond to the colors found by
\citet{Hadfield07} for Galactic WR stars; the solid line shows a power
law model F$_\nu \propto \nu^{\alpha}$, for $0\leq\alpha<2$. The
difference in colors between the LMC WR stars and the Galactic WR stars
is a combination of extinction, as illustrated by the reddening vectors,
and the large fraction of dusty WC9 stars among Galactic WR stars.}
\label{wrjkk80}
\end{figure}

\begin{figure}[ht]
\epsscale{1.00}
\plottwo{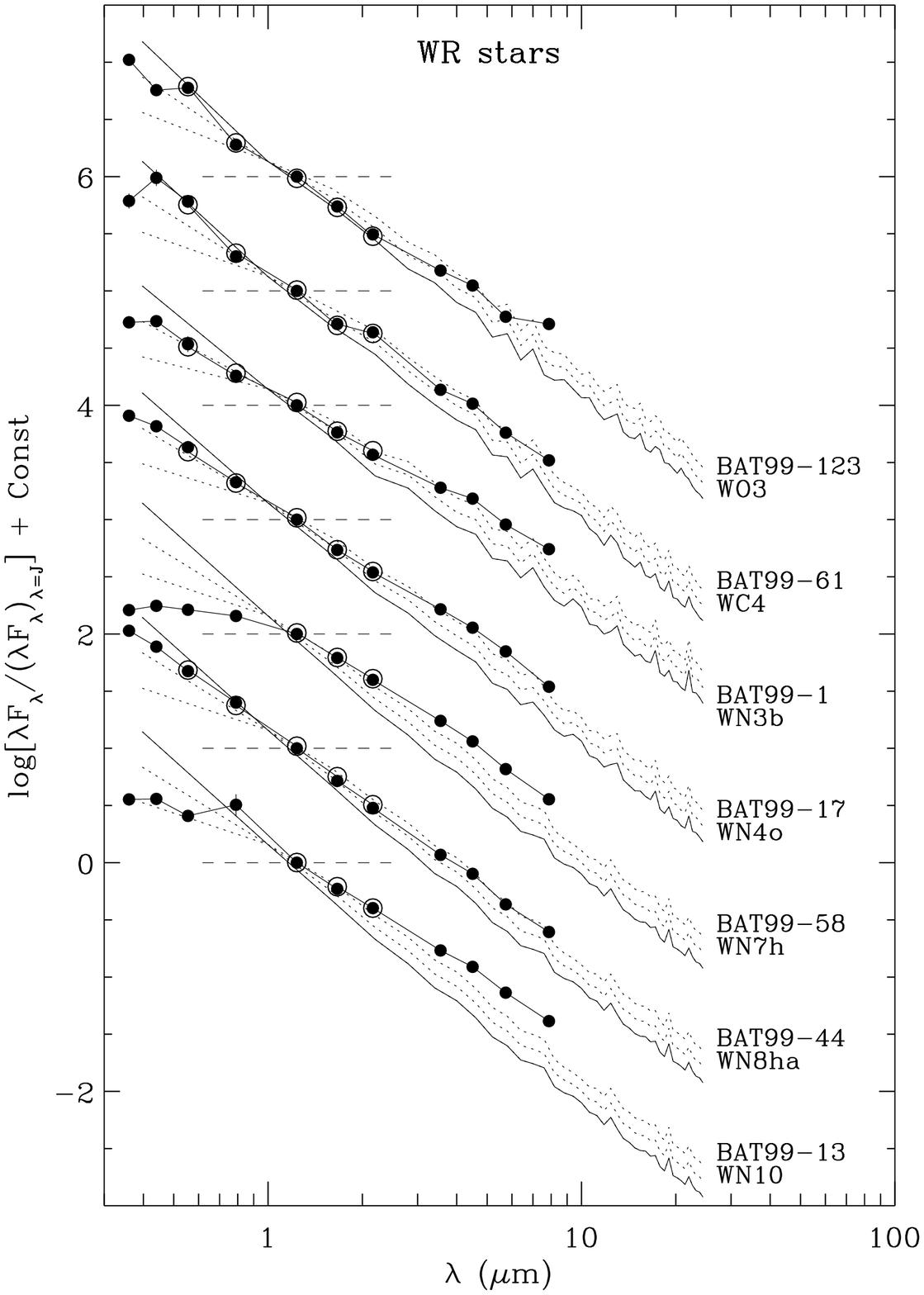}{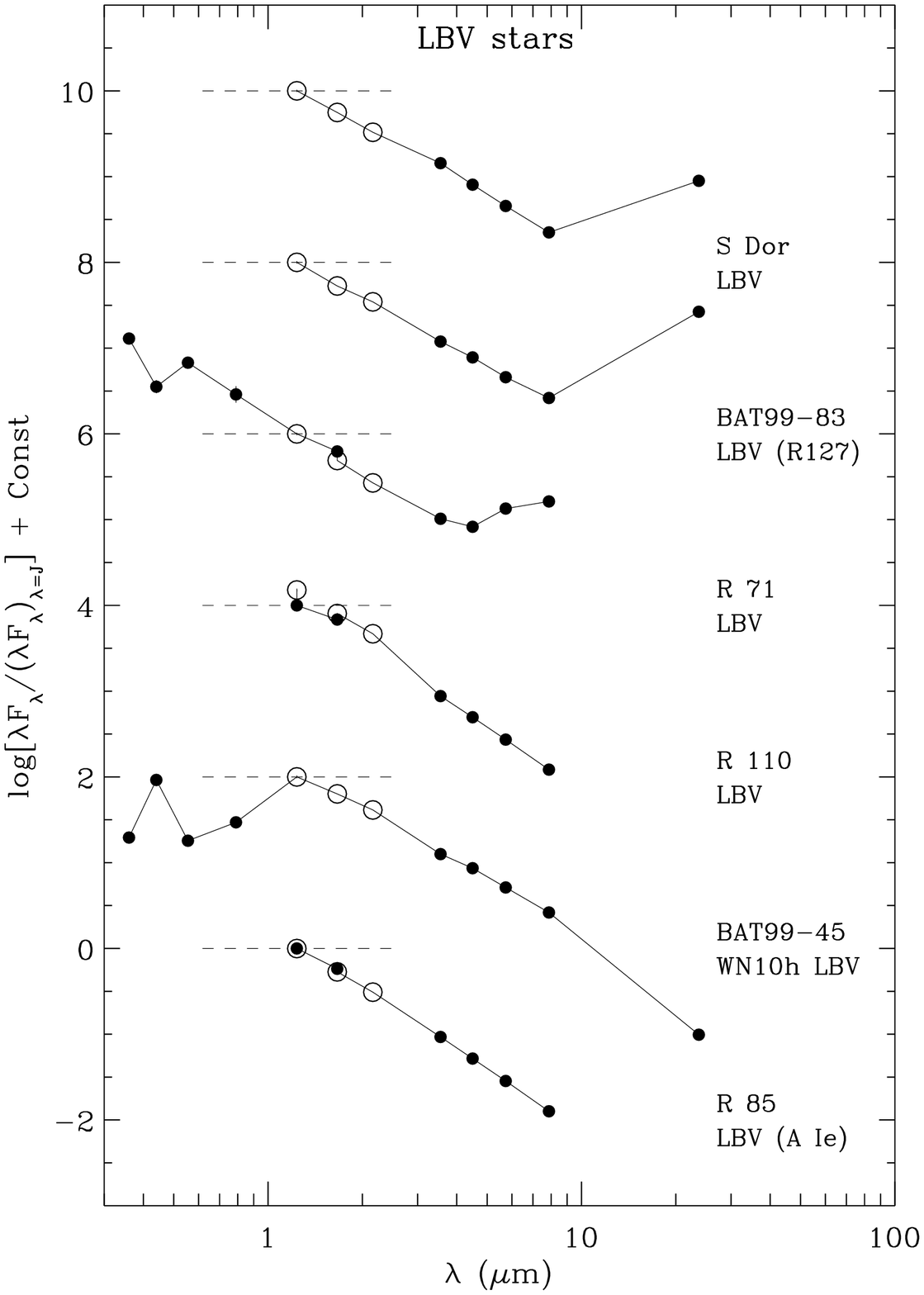}
\caption{Same as Figure~\ref{fig:obstack}, but for Wolf-Rayet stars
$(Left)$ and Luminous Blue Variables $(Right)$. For the WR stars, we
show model fits for BAT99-123, BAT99-61 and BAT99-17 and generic CMFGEN
models for the rest (see \S\ref{sec:wr}). All show excess above that
predicted by the models. The first 3 LBVs show evidence for dust;
BAT99-83 and R71 (saturated at $[24]$) were detected in the MIPS70
band. The different shapes of the LBV SEDs are likely related to the
time since the last outburst event and the amount of dust formed.
\label{fig:wr-lbvstack}}
\end{figure}

\begin{figure}[ht]
\epsscale{1.00}
\plottwo{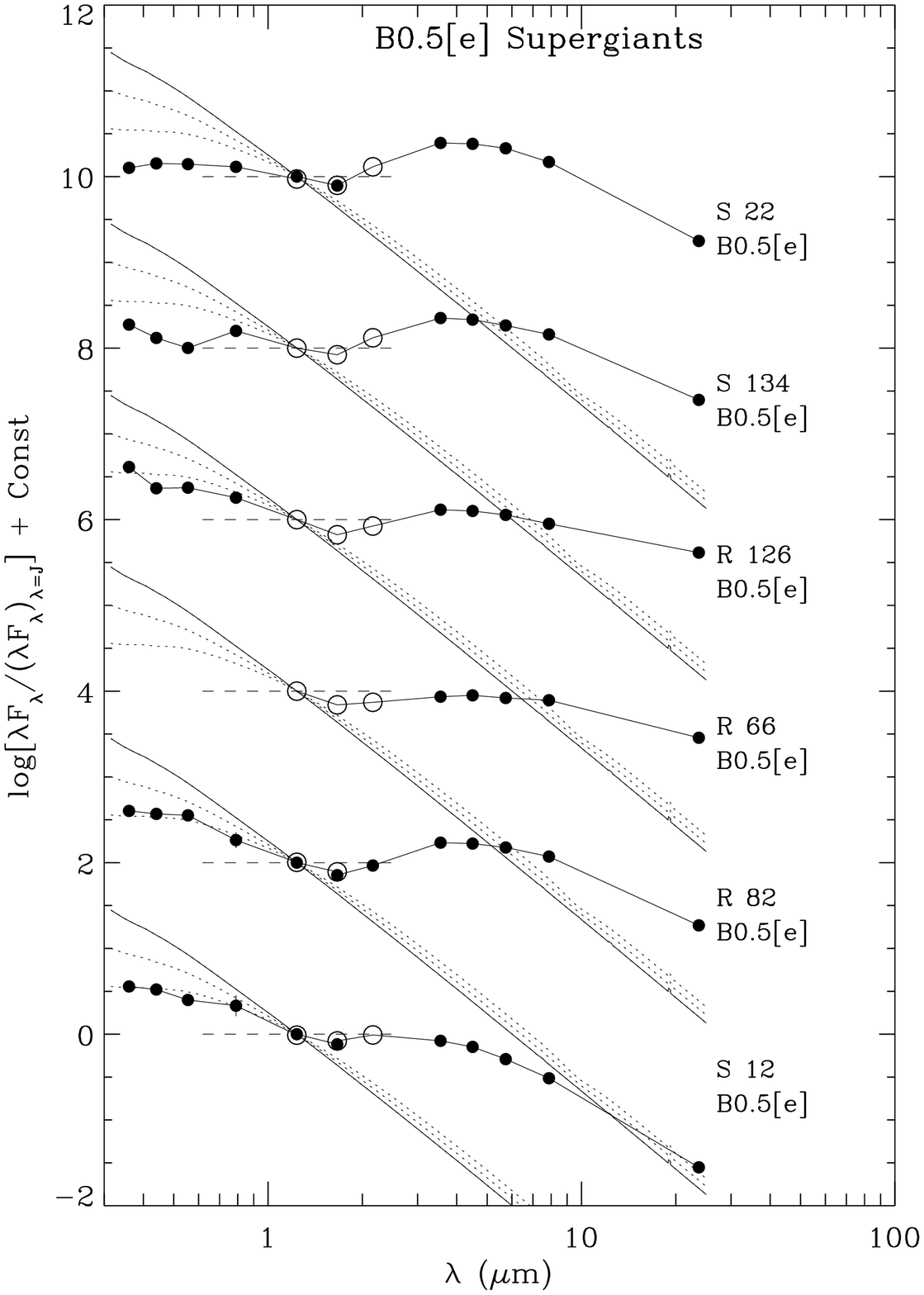}{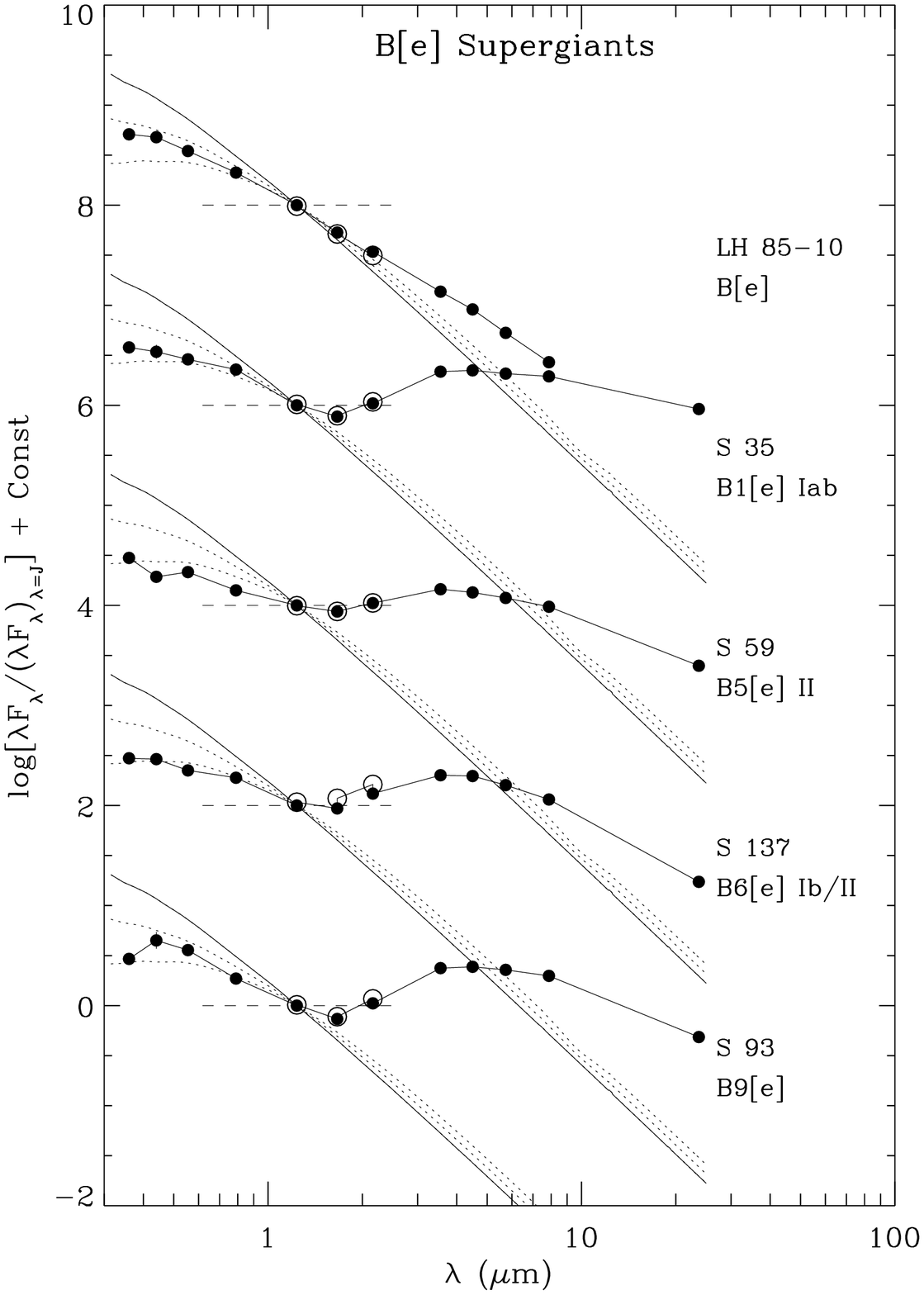}
\caption{Same as Figure~\ref{fig:obstack}, but for B[e] supergiants,
with a 20kK, $\log g = 2.25$ TLUSTY model overplotted. The sgB[e] have
similar SEDs revealing dust, except for LH~85-10, which has a SED
similar to that of a Be star. R126, R66, and S35 were also detected in
the MIPS70 band (see \S\ref{sec:sgBe}).
\label{fig:sgbestack}}
\end{figure}

\begin{figure}[ht]
\includegraphics[width=6in]{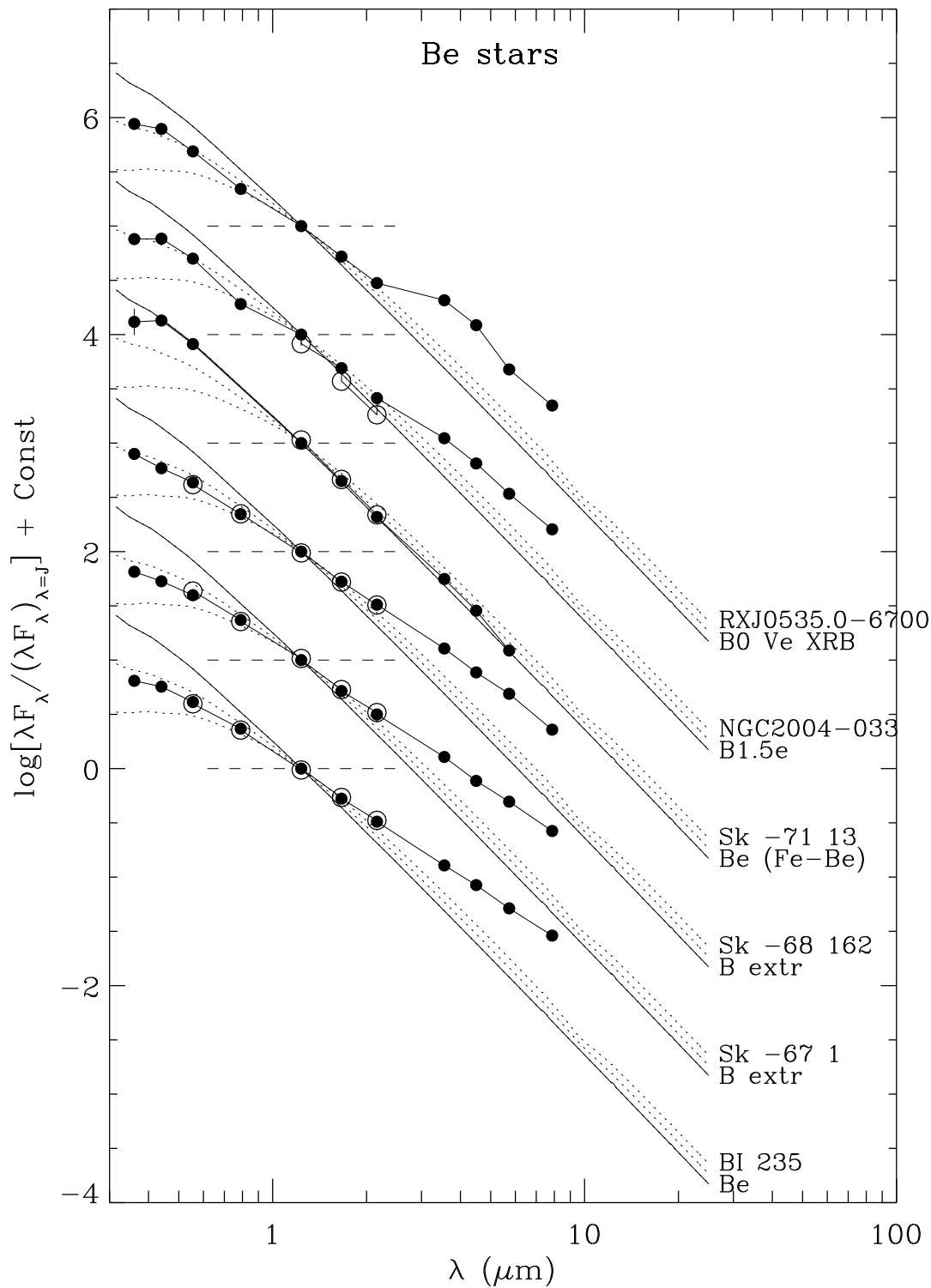}
\caption{Same as Figure~\ref{fig:obstack}, but for classical Be stars,
with a 25kK, $\log g = 2.75$ TLUSTY model overplotted. The similarity of
their SEDs implies that the various spectral types refer to the same
type of object. The Be/X-ray binary RXJ0535.0-6700 exhibits excess
emission at [3.6] and [4.5].
\label{fig:bestack}}
\end{figure}

The red K and M supergiants are among the brightest stars in the LMC at
all infrared wavelengths. However at 24~$\mu$m they span 5.5 magnitudes
(see Figure~\ref{cmd24}). This is due to the different types and amounts
of dust contributing to the SED at these wavelengths due to different
mass-loss rates \citep[see][]{Verhoelst09}. \citet{Josselin00} note that
the $K-[12]$ color index is a mass-loss indicator; therefore, we
similarly interpret the range in [24] magnitudes and the 2 magnitude
spread in $K_s-[8.0]$ color in Figure~\ref{jkk80} as a reflection of the
range of mass-loss rates. In Figure~\ref{iraccc2}, we notice a
temperature trend: red supergiants clustered at the bottom left
(i.e. with ``blue'' colors) are predominantly K or early M supergiants,
while the bulk of M0--M4 supergiants are found in a band, roughly
following the RSG model described in \S\ref{sec:colorcolor}. The
temperature trend correlates with the optical depth of the dust emission
and the amount of dust, which in turn is a function of the
mass-loss. The offset from the blackbody model can be explained by dust;
exceptions to this trend might be due to misclassifications, blends, or
variability. The same temperature trend is seen in Figure~\ref{cmd24},
with the coolest RSG being the most luminous at 24~$\mu$m. Our catalog
includes LMC 170452, which is a rare example of a RSG that has changed
spectral types. \citet{Levesque07} found it to vary from M4.5-M5 I in
2001 and 2004 to M1.5 I in 2005 December. The SAGE observations were
also taken in 2005 (July \& October), possibly during the transition. We
suggest that supergiants not appearing to follow the temperature trend
in Figure~\ref{iraccc2} could signify spectral variability (e.g. the
``blue'' M2 star [SP77]55-20 and the very red K0 star 165543). The
extremely reddened (at long wavelengths) outlier in Figures~\ref{cmd8},
\ref{cmd24}, and \ref{iraccc2} is the M2-3 supergiant 175549 with
$[8]-[24]=4$~mag. The M0~I 138475 has $K_s-[8.0]=3$~mag, making it an
outlier in Figure~\ref{jkk80}. In Figure~\ref{cmd36j36}, it is located
with the AGB stars, implying it is an M-type AGB star.

\begin{figure}[ht]
\epsscale{1.00}
\plottwo{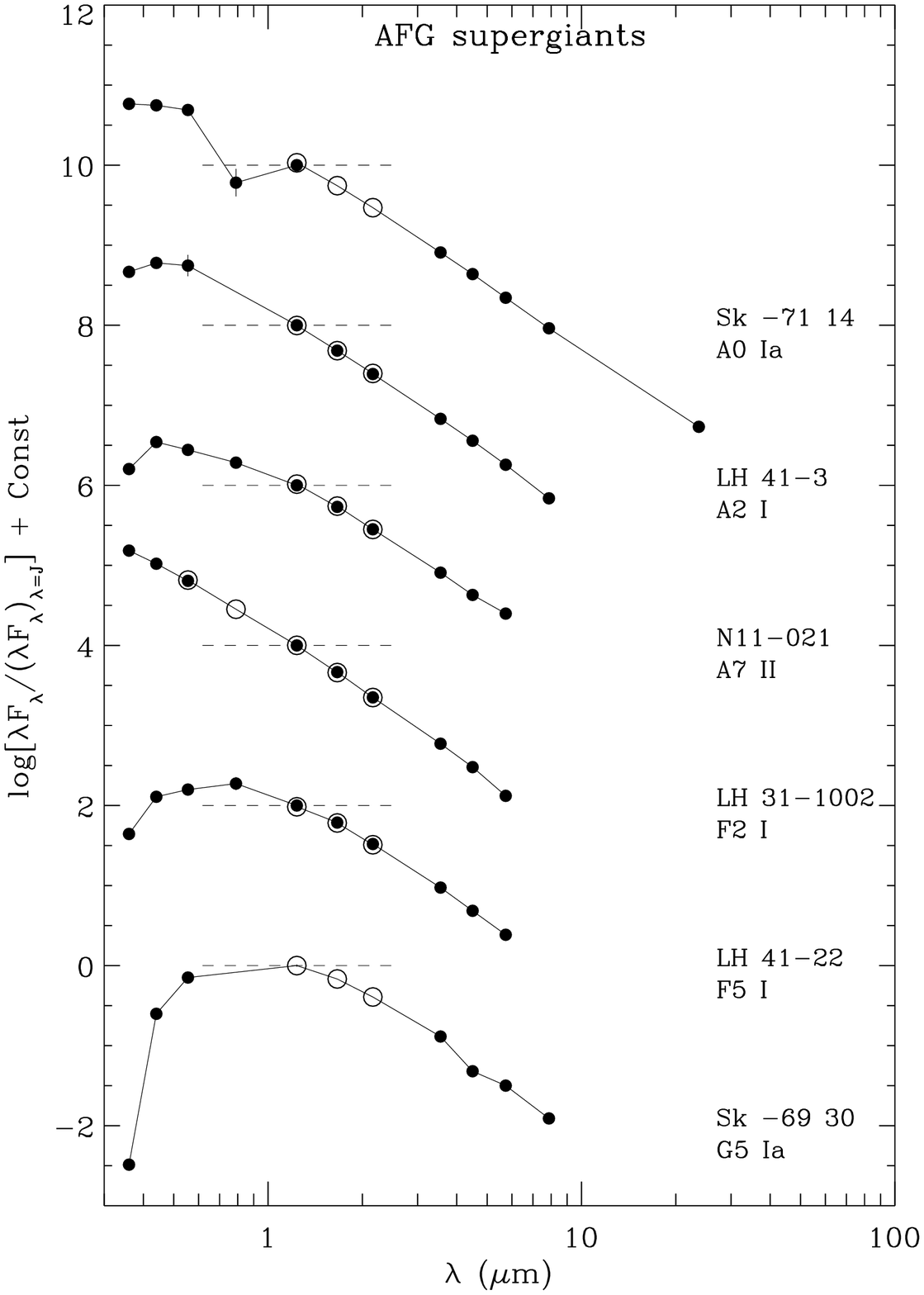}{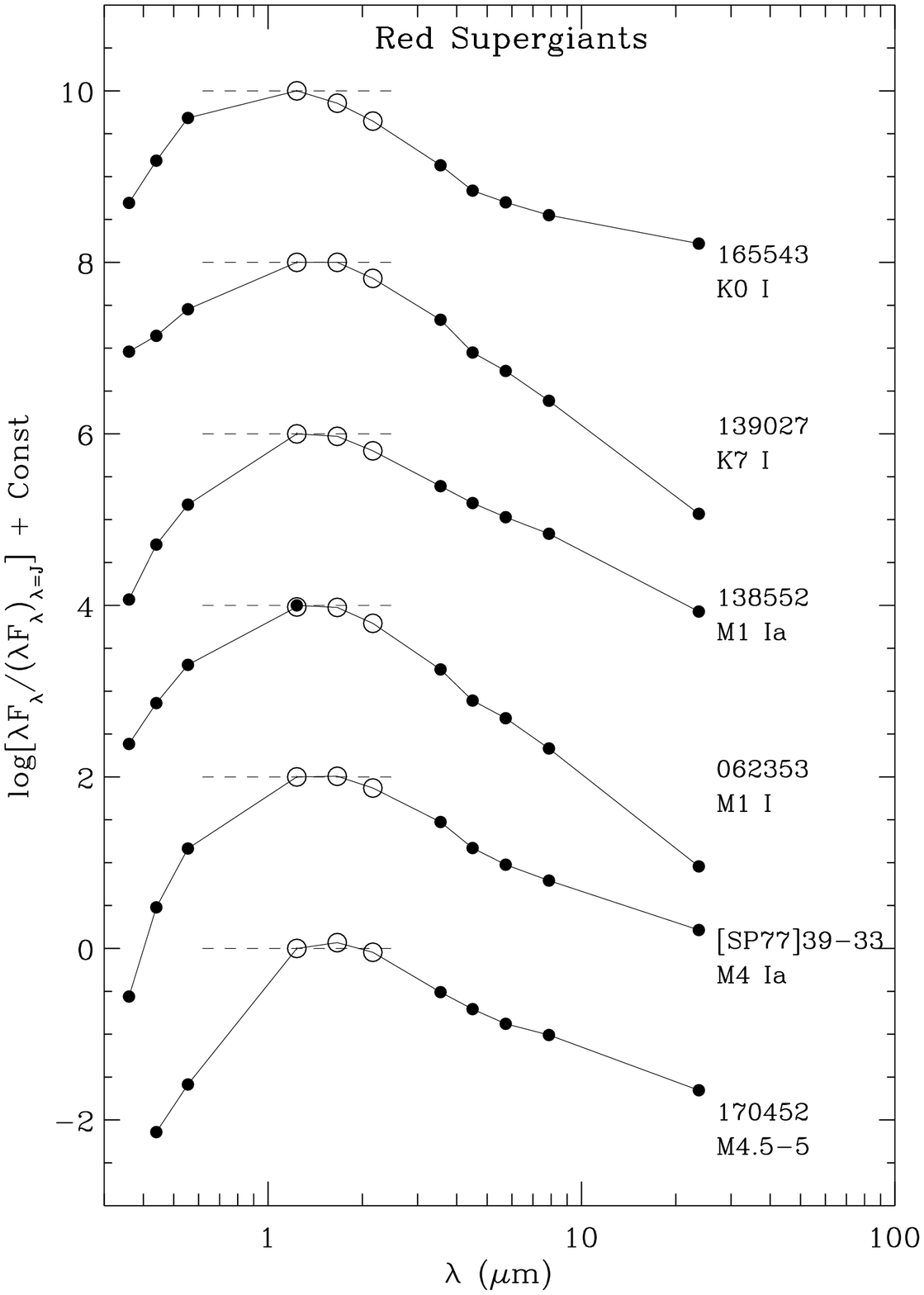}
\caption{Same as Figure~\ref{fig:obstack}, but for A, F and G
supergiants $(Left)$ and K and M supergiants $(Right)$. No reference
models are shown. The SED of LH~31-1002 (F2~I) implies a hot effective
temperature, inconsistent with its spectral type. The K7~I star 139027
has an excess at the shortest wavelengths, possibly indicating a hot
companion. The M4.5-5 supergiant 170452 is a rare case of a RSG that
changed spectral type (see \S\ref{sec:other}).
\label{fig:a-mstack}}
\end{figure}

Figure~\ref{fig:a-mstack} shows representative SEDs of yellow and red
supergiants. The SED peak moves redward with decreasing temperature, as
expected. Note that starting at Sk~$-69^\circ$~30 (G5 I), a depression
in the continuum due to the CO band appears at 4.5~$\mu$m, and persists
to later types. This depression is extremely weak or absent in the M1
Ia, 138552. Some of the later supergiants have a distinct break in their
slopes at roughly 4.5~$\mu$m, also coinciding with the CO band. The SED
of LH~31-1002 (F2~I), suggests a hotter effective temperature,
inconsistent with its spectral type. \citet{Massey00} give $B-V=0.4$ mag
for this star, in disagreement with the value from MCPS ($B-V=-0.2$
mag), perhaps suggesting an error in its coordinates that lead to a
match with a nearby hot star. The K7~I star 139027 also appears to have
an excess at the shortest wavelengths; therefore it is a candidate
$\zeta$ Aurigae binary \citep{Wright70}, with a hot companion. LMC
170452, has a peculiar SED (with $V-K=7.2$ mag, instead of $\sim5$),
possibly reflecting its spectral type change.

\begin{figure}[ht]
\includegraphics[width=6in]{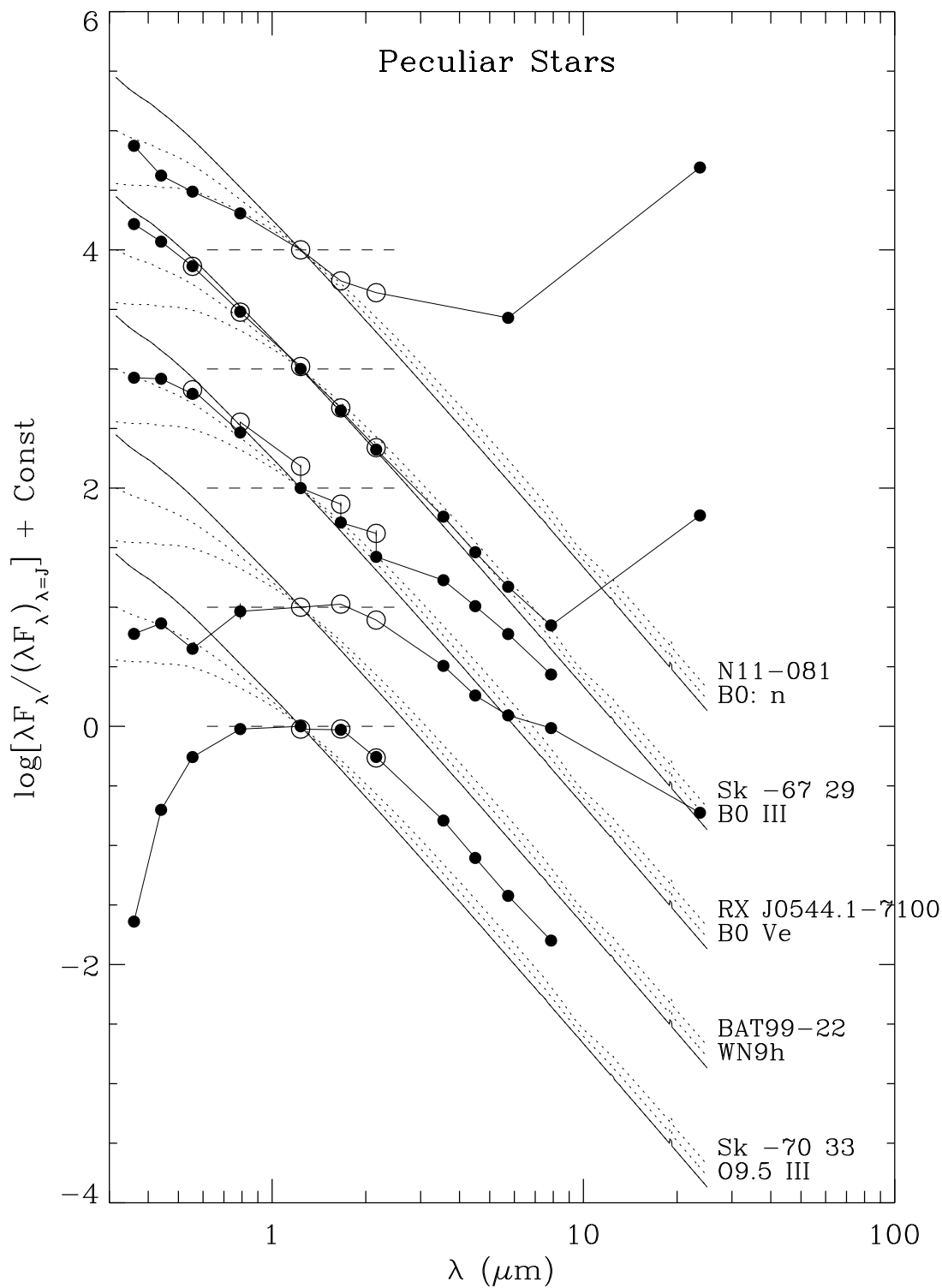}
\caption{Same as Figure~\ref{fig:obstack}, but for peculiar SEDs,
arising from nebular contamination (N11-081), variability
(RXJ0544.1-7100), blending (BAT99-22) or misidentification at one
(Sk~$-67^\circ$~29; at 24~$\mu$m), or all (Sk~$-70^\circ$~33) bands.
\label{fig:pecstack}}
\end{figure}

Finally, in Figure~\ref{fig:pecstack} we present some peculiar SEDs,
illustrating examples of nebular contamination (N11-081), variability
(RXJ0544.1-7100), blending (BAT99-22) or misidentification
(Sk~$-67^\circ$~29, Sk~$-70^\circ$~33) in our photometric
catalog. Nebular contamination and misidentifications are responsible
for the OB outliers in the CMDs and TCDs. Only 3 out of 354 O stars
(N11-028, N11-046, lmc2-703-- all main-sequence stars) and 15 out of 586
early-B stars have 24~$\mu$m detections, illustrating that contamination
is less than 3\% at this wavelength.

\section{Summary}
\label{sec:summary}

This paper presents the first major catalogs of accurate spectral types
and multi-wavelength photometry of massive stars in the LMC. The
spectroscopic catalog contains 1750 massive stars, with accurate
positions and spectral types compiled from the literature, and includes
the largest modern spectroscopic surveys of hot massive stars in the
LMC. The photometric catalog comprises uniform $0.3-24$ $\mu$m
photometry in the $UBVIJHK_{s}$+IRAC+MIPS24 bands for a subset of 1268
stars that were matched in the SAGE database. Our photometric catalog
increases the infrared photometry for OB stars presented in a single
study by an order of magnitude, allowing for a detailed study of their
wind parameters. The low foreground reddening toward the LMC and the
identical distance of the stars in the sample remove degeneracies
inherent in Galactic studies and enable the investigation of infrared
excesses, while minimizing systematic errors due to reddening.

We examine the infrared excesses of the half-solar metallicity OB stars
in the LMC, finding a correlation of excess with wavelength, as
expected, and luminosity. The dispersion in the relation reflects the
range of mechanisms, and therefore stellar wind parameters, which
produce free--free emission in OB stars. We discuss the positions of OB
stars, WR, LBV, sgB[e], classical Be stars, RSG, AFG supergiants and
Be/X-ray binaries on CMDs and TCDs, which serve as a roadmap for
interpreting existing infrared photometry of massive stars in nearby
galaxies. We conclude that sgB[e], RSG and LBVs are among the most
luminous stars in the LMC at all infrared wavelengths. Representative
SEDs are shown for all types of stars, and in certain cases are compared
to model atmospheres to illustrate infrared excesses due to free--free
emission from winds or disks. We confirm the presence of dust around 10
sgB[e] stars from the shape of their spectral energy distributions,
which are presented for the first time at these mid-infrared
wavelengths, and find the SED shapes to be very similar. The large
luminosities and the presence of dust characterizing both B[e]
supergiants ($\log L/L_{\odot}\geq4$) and the rare, dusty progenitors of
the new class of optical transients (e.g. SN~2008S and NGC~300~OT),
suggest a common origin for these objects. The variety of SED shapes
observed among the LBVs is likely related to the time since the last
outburst event and the amount of dust formed. Finally, we find the
distribution of infrared colors for WR stars to differ from that of
Galactic WR stars, due to the abundance of Galactic dusty WC9 types and
the effects of foreground reddening in the Milky Way.

This work demonstrates the wealth of information contained in the SAGE
survey and aims to motivate more detailed studies of the massive stars
in the LMC.  Examples of studies that can be conducted using our
catalogs follow: (a) Candidate RSG, sgB[e], LBV in the LMC and other
galaxies can be identified by selecting them from infrared CMDs, ruling
out AGB stars from optical and near-infrared photometry (if it exists),
and confirming these with spectroscopy. For example, in IC 1613, a
comparison of the IRAC CMD in Figure~4 of \citet{Jackson07b} to our
Figure~\ref{cmd36}, suggests that the brightest ``blue'' stars are
candidate RSG or LBV stars. With the optical catalog of
\citet{Garcia09}, their visual colors can be determined and 0.3-8~$\mu$m
SEDs can be constructed. (b) Specific stars or categories of stars can
be selected from the catalog for further study. For example, mass-loss
rates can be determined for OB stars that have spectra from the {\it Far
Ultraviolet Spectroscopic Explorer} \citep[e.g.][]{Fullerton00}, since
terminal velocities can be measured from the ultraviolet lines. (c) The
OB stars can be used to determine the LMC reddening law, given the known
spectral types and SED shapes of these stars. (d) Variability in
2MASS--IRSF photometry can be used to pick out new, variable, massive
stars. The most luminous stars in our catalog ($V\sim10$~mag) are
saturated in the LMC microlensing surveys; however, the two epochs of
near-infrared photometry can be used, for example, to find candidate
eclipsing binaries not included in the OGLE \citep{Wyrzykowski03} or
MACHO \citep{Faccioli07} lists. (e) Our catalog can be cross correlated
with X-ray catalogs of the LMC to study non-thermal emission processes
in massive stars.

A comparison of the infrared properties of massive stars in the Large
vs. Small Magellanic Cloud (SMC) will be pursued next. Stellar winds
scale with metallicity \citep{Mokiem07b}; therefore, a comparison of
infrared excesses of OB stars in the LMC vs. the SMC will help quantify
the effect of metallicity on OB star winds, Be star properties, sgB[e]
stars etc. In the future, the {\it James Webb Space Telescope} will
obtain infrared photometry of massive stars in galaxies out to 10 Mpc
and measuring the colors of massive stars as a function of metallicity,
will be useful for interpreting these data.

\acknowledgments{We acknowledge the input from members of the Massive
  Stars \& Starbursts research group at STScI, in particular, Nolan
  Walborn. We thank Bernie Shao for incorporating the OGLE~III
  photometry to the database and Paul Crowther for providing us with his
  CMFGEN models of WR stars. AZB acknowledges support from the Riccardo
  Giacconi Fellowship award of the Space Telescope Science
  Institute. The {\it Spitzer} SAGE project was supported by NASA/{\it
    Spitzer} grant 1275598 and NASA NAG5-12595. This work is based [in
    part] on archival data obtained with the {\it Spitzer} Space
  Telescope, which is operated by the Jet Propulsion Laboratory,
  California Institute of Technology under a contract with NASA. Support
  for this work was provided by an award issued by JPL/Caltech. This
  publication makes use of data products from the Two Micron All Sky
  Survey, which is a joint project of the University of Massachusetts
  and the Infrared Processing and Analysis Center/California Institute
  of Technology, funded by the National Aeronautics and Space
  Administration and the National Science Foundation.}

{\it Facility:} \facility{Spitzer (IRAC, MIPS)}

\clearpage
\section*{Appendix}

We adopted a reddening curve composed of a continuous contribution and a
feature which represents silicate absorption at 9.7~$\mu$m.  The
continuous portion is based on the formulation used by
\citet{Fitzpatrick09} to fit near-infrared extinction curves (which
include 2MASS $JHK_s$) and is a generalization of the analytic formula
given by \citet{Pei92}. For $\lambda\geq\lambda_0$, it has the form
\begin{equation}
k(\lambda -V)_c \equiv \frac{E(\lambda-V)}{E(B-V)} = \frac{0.349+ 
         2.087 R(V)}{1  +(\lambda/\lambda_0)^\alpha} -R(V),
\end{equation}
where $\lambda_0=0.507\,\mu$m. The ratio of total to selective
extinction, $R(V) \equiv A(V)/E(B-V)$, and $\alpha$\ are free
parameters.  Near-infrared observations show that these parameters can
vary from one sight line to the next.  In this work, we use a curve
defined by $\alpha = 2.05$ and $R(V) = 3.11$.  These parameters provide
a good representation of the mean curve determined by
\citet{Fitzpatrick09} and also to extinction data taken from the
literature for diffuse interstellar medium sight lines in the Milky Way.
The latter data sets extend to 24~$\mu$m and include the following:
\citet{Castor83}, \citet{Koornneef83}, \citet{Abbott84},
\citet{Leitherer84}, \citet{Rieke89}, \citet{Wegner94}, and
\citet{Nishiyama09}. Although we cannot be certain this same form
applies to the LMC, it is consistent with the SEDs of lightly reddened,
normal hot stars that are thought to have weak winds (see
\S\ref{sec:ob}).  In any case, the extinction for most of our objects is
small enough that inaccuracies in the interpretation of reddening should
be minimal. For completeness, we included a Drude profile \citep[see,
e.g.,][]{Fitzpatrick09} to describe the 9.7 $\mu$m\ silicate feature,
even though it has little effect on the {\it Spitzer} bands.  The
adopted profile is given by
\begin{equation}
f(\lambda) = \frac{a_3 x^2}{ (x^2 -x_0^2)^2 +(x\gamma)^2},
\end{equation}
where $x = 1/\lambda \; \mu{\mbox m}^{-1}$, and the parameters were
assigned the values $x_0 = 1/9.7 \; \mu{\mbox m}^{-1}$, $\gamma = 0.03$
and $a_3 = 3 \times 10^{-4}$. Finally, the complete extinction curve is
given by $k(\lambda -V) = k(\lambda -V)_c + f(\lambda)$.  When we plot
the SEDs, we actually use the form $k(\lambda -J) = k(\lambda -V)
-k(J-V)$.

\clearpage

\clearpage

\clearpage

\clearpage

\clearpage


\input{tab1.tex}
\input{tab2.tex}
\input{tab3stub.tex}
\input{tab4.tex}
\input{tab5.tex}

\end{document}

%% file: tab1.tex
\begin{deluxetable}{lcccl}
\tabletypesize{\footnotesize}
\tablewidth{0pc}
\tablecaption{\sc Catalog of Spectral Types for 1750 LMC Massive Stars}
\tablehead{\colhead{Star} & \colhead{RA (J2000)} &
\colhead{Dec (J2000)} & \colhead{Reference$^{b}$} & \colhead{Classification} \\
\colhead{Name$^{a}$} & \colhead{(deg)} &
\colhead{(deg)} & \colhead{} & \colhead{\& Comments}}
\startdata
BAT99$-$1  & 71.3843765 &  $-$70.2530289 &  F03 &       WN3b Sk $-$70 1          \\
  Sk $-$69 1  & 71.3865433 &  $-$69.4670029 &  C86 &         B1 III               \\
  Sk $-$67 1  & 71.4227524 &  $-$67.5705261 &  C86 &          B extr               \\
  Sk $-$67 2  & 71.7685394 &  $-$67.1147461 &  F88 &         B1 Ia+               \\
 Sk $-$70 1a  & 71.8649979 &  $-$70.5626984 &  C86 &         O9  II               \\
  Sk $-$69 8  & 72.3293762 &  $-$69.4318314 &  F88 &         B5  Ia               \\
BAT99$-$2  & 72.4009000 &  $-$69.3484000 &  F03 &    WN2b(h)                   \\
  Sk $-$69 9  & 72.4642105 &  $-$69.2011948 &  J01 &       O6.5 III               \\
\enddata                         
\label{tab:catalog}
\tablenotetext{a}{Star designations: \citet[BAT99;][]{Breysacher99},
\citet[Sk;][]{Sanduleak70}, \citet[W;][]{Westerlund61},
\citet[BI;][]{Brunet75}, \citet[LH;][]{Lucke72}, \citet[S;][]{Henize56}}
\tablenotetext{b}{Reference key: (B01) \citet{Bartzakos01}, (B09)
\citet{Bonanos09}, (C86) \citet{Conti86}, (E06) \citet{Evans06}, (F88)
\citet{Fitzpatrick88, Fitzpatrick91}, (F02) \citet{Fitzpatrick02},
(F03a) \citet{Fitzpatrick03}, (F03) \citet{Foellmi03}, (G94)
\citet{Garmany94}, (G05) \citet{Gonzalez05}, (G98) \citet{Guinan98},
(H92) \citet{Heydari92}, (H03) \citet{Heydari03}, (H79)
\citet{Humphreys79}, (H94) \citet{Humphreys94}, (J01) \citet{Jaxon01},
(L05) \citet{Liu05}, (M95) \citet{Massey95}, (M00) \citet{Massey00},
(M02) \citet{Massey02}, (M02b) \citet{Massey02b}, (M03)
\citet{Massey03}, (M04) \citet{Massey04}, (M05) \citet{Massey05}, (M09)
\citet{Massey09}, (Me05) \citet{Meynadier05}, (N94) \citet{Niemela94},
(N01) \citet{Niemela01}, (O95) \citet{Oey95}, (O01) \citet{Olsen01},
(O01b) \citet{Ostrov01b}, (O01c) \citet{Ostrov01c}, (O03)
\citet{Ostrov03}, (P01) \citet{Parker01}, (P92) \citet{Parker92}, (P93)
\citet{Parker93}, (R02) \citet{Ribas02}, (R05) \citet{Raguzova05}, (S92)
\citet{Schild92}, (S08) \citet{Schnurr08}, (T98) \citet{Testor98}, (W77)
\citet{Walborn77}, (W97) \citet{Walborn97}, (W02) \citet{Walborn02},
(W03) \citet{Walborn03}, (W04) \citet{Walborn04}, (W08)
\citet{Williams08}, (Z06) \citet{Zickgraf06}.}
\tablecomments{Table~\ref{tab:catalog} is available in its entirety in the
electronic version of the Journal. A portion is shown here for guidance
regarding its form and content.}
\end{deluxetable} 

%% file: tab2.tex
\begin{deluxetable}{ll}
\tabletypesize{\footnotesize}
\tablewidth{0pc}
\tablecaption{\sc Statistics for the 1268 Matched Stars}
\tablehead{\colhead{Catalogs Matched} & \colhead{\# Stars} }

\startdata
IRACC	                    &    5  \\
IRACC+IRSF	            &   88 \\
IRACC+IRSF+MCPS	            &  601  \\
IRACC+IRSF+MCPS+MIPS24	    &  122 \\
IRACC+IRSF+MCPS+MIPS24+OGLE &  9 \\
IRACC+IRSF+MCPS+OGLE	    &  364 \\
IRACC+IRSF+MIPS24	    &   21 \\
IRACC+IRSF+OGLE	            &   18 \\
IRACC+MCPS	            &   33 \\
IRACC+MCPS+MIPS24	    &    1 \\
\tableline
IRACA+MIPS24                 &   1 \\
IRACA+MCPS+MIPS24            &   1 \\
IRACCEP1+MCPS+MIPS24         &   4 \\
\enddata                         
\label{tab:matchtype}
\end{deluxetable} 

%% file: tab3stub.tex
\begin{deluxetable}{lllllllllll}
\rotate
\tabletypesize{\footnotesize}
\tablewidth{0pc}
\tablecaption{\sc 0.3-24 $\mu${\rm m} Photometry of 1268 Massive Stars in the LMC}
\tablehead{\colhead{Star Name$^a$} & \colhead{IRAC Designation} &
\colhead{RA(J2000)} & \colhead{Dec(J2000)} &
\colhead{$U$} & \colhead{$\sigma_U$} &
\colhead{$B$} & \colhead{$\sigma_B$} &\colhead{...} &
\colhead{Ref.$^b$} & \colhead{Classification \& Comments}}
\startdata
BAT99$-$1  & J044532.25$-$701510.8 & 71.3843765 & $-$70.2530289 & 15.001 & 0.037 & 15.647 & 0.029 & ...& F03 & WN3b Sk $-$70 1 \\
Sk $-$69 1  & J044532.76$-$692801.2 & 71.3865433 & $-$69.4670029 & 12.343 & 0.01 & 13.364 & 0.01 &...&  C86 & B1 III \\
Sk $-$67 1  & J044541.44$-$673413.9 & 71.4227524 & $-$67.5705261 & 12.829 & 0.038 & 13.72 & 0.025 &...&  C86 & B extr  \\
Sk $-$67 2  & J044704.44$-$670653.1 & 71.7685394 & $-$67.1147461 & 10.351 & 0.104 & 11.219 & 0.095 &...& F88 & B1 Ia+  \\
Sk $-$70 1a  & J044727.58$-$703345.6 & 71.8649979 & $-$70.5626984 & 12.693 & 0.01 & 13.654 & 0.01 &...&  C86 & O9 II \\
Sk $-$69 8  & J044919.03$-$692554.6 & 72.3293762 & $-$69.4318314 & 10.838 & 0.06 & 11.714 & 0.078 & ...& F88 & B5 Ia \\
BAT99$-$2  & J044936.24$-$692054.8 & 72.4009 & $-$69.3484 & 15.633 & 0.038 & 16.495 & 0.03 &...& F03 & WN2b(h) \\
\enddata                         
\label{tab:phot}
\tablenotetext{a}{Star designations: \citet[BAT99;][]{Breysacher99},
\citet[Sk;][]{Sanduleak70}, \citet[W;][]{Westerlund61},
\citet[BI;][]{Brunet75}, \citet[LH;][]{Lucke72}, \citet[S;][]{Henize56}}
\tablenotetext{b}{Reference key: (B01) \citet{Bartzakos01}, (B09)
\citet{Bonanos09}, (C86) \citet{Conti86}, (E06) \citet{Evans06}, (F88)
\citet{Fitzpatrick88, Fitzpatrick91}, (F02) \citet{Fitzpatrick02},
(F03a) \citet{Fitzpatrick03}, (F03) \citet{Foellmi03}, (G94)
\citet{Garmany94}, (G05) \citet{Gonzalez05}, (G98) \citet{Guinan98},
(H92) \citet{Heydari92}, (H03) \citet{Heydari03}, (H79)
\citet{Humphreys79}, (H94) \citet{Humphreys94}, (J01) \citet{Jaxon01},
(L05) \citet{Liu05}, (M95) \citet{Massey95}, (M00) \citet{Massey00},
(M02) \citet{Massey02}, (M02b) \citet{Massey02b}, (M03)
\citet{Massey03}, (M04) \citet{Massey04}, (M05) \citet{Massey05}, (M09)
\citet{Massey09}, (Me05) \citet{Meynadier05}, (N94) \citet{Niemela94},
(N01) \citet{Niemela01}, (O95) \citet{Oey95}, (O01) \citet{Olsen01},
(O01b) \citet{Ostrov01b}, (O01c) \citet{Ostrov01c}, (O03)
\citet{Ostrov03}, (P01) \citet{Parker01}, (P92) \citet{Parker92}, (P93)
\citet{Parker93}, (R02) \citet{Ribas02}, (R05) \citet{Raguzova05}, (S92)
\citet{Schild92}, (S08) \citet{Schnurr08}, (T98) \citet{Testor98}, (W77)
\citet{Walborn77}, (W97) \citet{Walborn97}, (W02) \citet{Walborn02},
(W03) \citet{Walborn03}, (W04) \citet{Walborn04}, (W08)
\citet{Williams08}, (Z06) \citet{Zickgraf06}.}
\tablecomments{Table~\ref{tab:phot} is available in its entirety in the
electronic version of the Journal. A portion is shown here for guidance
regarding its form and content.}
\end{deluxetable} 

%% file: tab4.tex
\begin{deluxetable}{lcccc}
\tabletypesize{\footnotesize}
\tablewidth{0pc}
\tablecaption{\sc Filter \& Detection Characteristics}
\tablehead{\colhead{Filter}  & \colhead{$\lambda_{\rm eff}$}  &
\colhead{Zero mag} & \colhead{Resolution} & \colhead{\# stars}\\
\colhead{}  & \colhead{($\mu$m)}  &
\colhead{flux ($Jy$)} & \colhead{($\arcsec$)} & \colhead{detected}}
\startdata
$U$  &        0.36  &    1790 & 1.5/2.6 &  1121   \\
$B$ & 	      0.44  &    4063 & 1.5/2.6 &  1136   \\
$V$ & 	      0.555  &    3636 & 1.5/2.6 &  1136   \\
$I$ & 	      0.79  &    2416 & 1.5/2.6 &  1026   \\
$V_{OGLE}$ &  0.555  &    3636 & 1.2 &   391   \\
$I_{OGLE}$ &  0.79  &    2416 & 1.2 &   391   \\
$J$ & 	      1.235   &   1594  & 2.5 &  1203   \\
$H$ & 	      1.662   &   1024  & 2.5 &  1218   \\
$K_s$ &       2.159   &   666.7  & 2.5 &  1184   \\
$J_{IRSF}$ &  1.235   &   1594  & 1.3 &  1122   \\
$H_{IRSF}$ &  1.662   &   1024  & 1.3 &  1089   \\
$Ks_{IRSF}$ & 2.159   &   666.7  & 1.3 &  1077   \\
$[3.6]$ &     3.55   &   280.9  & 1.7 &   1260  \\
$[4.5]$ &     4.493   &   179.7  & 1.7 &   1234  \\
$[5.8]$ &     5.731   &   115.0  & 1.9 &    950  \\
$[8.0]$ &     7.872   &   64.13  & 2 &    577  \\
$[24]$ &      23.68   &   7.15  & 6 &   159   \\
\enddata                         
\label{tab:filters}

\end{deluxetable} 

%% file: tab5.tex
\begin{deluxetable}{llllllllll}
\rotate
\tabletypesize{\footnotesize}
\tablewidth{0pc}
\tablecaption{\sc MIPS70+160 Photometry}
\tablehead{\colhead{Star Name}  & \colhead{Spectral Type}  &
\colhead{mag$_{70}$} & \colhead{$\sigma_{mag70}$} &
\colhead{Flux$_{70}$} & \colhead{$\sigma_{Flux70}$} &
\colhead{mag$_{160}$} & \colhead{$\sigma_{mag160}$} &
\colhead{Flux$_{160}$} & \colhead{$\sigma_{Flux160}$}\\
\colhead{}  & \colhead{} &
\colhead{(mag)} & \colhead{(mag)} &
\colhead{(mJy)} & \colhead{(mJy)} &
\colhead{(mag)} & \colhead{(mag)} &
\colhead{(mJy)}& \colhead{(mJy)}}
\startdata
BAT99-8 & WC4 & 1.379   & 0.02796  &  218.4 & 5.623 & \nodata  & \nodata  & \nodata  & \nodata \\
BAT99-22 & WN9h (+RSG) & 2.327   & 0.04657  &  91.24 & 3.911 & \nodata	& \nodata  & \nodata  & \nodata \\
BAT99-32 & WN6(h) & 1.4     & 0.03326  &  214.4 & 6.564 & \nodata	& \nodata  & \nodata  & \nodata \\
BAT99-37 & WN3o & 1.503   & 0.02847  &  194.8 & 5.107 & \nodata	& \nodata  & \nodata  & \nodata \\
BAT99-38 & WC4+$[$O8I:$]$ & -0.8254 & 0.01853  &  1664  & 28.4  & -2.412   & 0.1344   & 1476     & 181.8 \\
BAT99-53 & WC4 (+OB) & 1.386	 & 0.03978  &  217.1 & 7.952 & \nodata	& \nodata  & \nodata  & \nodata \\
BAT99-55 & WN11h & 0.6481	 & 0.03829  &  428.3 & 15.1  & \nodata	& \nodata  & \nodata  & \nodata \\
BAT99-85 & WC4 (+OB) & -0.4681 & 0.03374  &  1197  & 37.19 & \nodata	& \nodata  & \nodata  & \nodata \\
BAT99-123 & WO3 & 1.782	 & 0.04259  &  150.7 & 5.909 & \nodata	& \nodata  & \nodata  & \nodata \\
BAT99-133 & WN11h & 0.3187  & 0.01856  &  580.1 & 9.914 & \nodata	& \nodata  & \nodata  & \nodata \\
R66    & B0.5[e] & 0.4201	 & 0.01875  &  528.4 & 9.126 & \nodata	& \nodata  & \nodata  & \nodata \\
S35    & B1[e] Iab & 2.104	 & 0.05634  &  112   & 5.806 & \nodata	& \nodata  & \nodata  & \nodata \\
R126   & B0.5[e] & 0.993	 & 0.03794  &  311.7 & 10.89 & \nodata	& \nodata  & \nodata  & \nodata \\
R71    & LBV & -0.9245 & 0.01292  &  1823  & 21.7  & \nodata	& \nodata  & \nodata  & \nodata \\
R127   & LBV & -0.1097 & 0.03093  &  860.7 & 24.51 & \nodata	& \nodata  & \nodata  & \nodata \\
\enddata                         
\label{tab:mips70}

\end{deluxetable}